\documentclass[
    prd,
    amsfonts,
    onecolumn,
    superscriptaddress,
    aps,
    nofootinbib,
    11pt
]{revtex4-2}

\usepackage[utf8]{inputenc} 
\usepackage[T1]{fontenc}
\usepackage{microtype} 
\usepackage{ragged2e} 
\usepackage{xcolor}

\usepackage{amsmath, amssymb, amsfonts}

\usepackage[dvipsnames]{xcolor} 

\usepackage{graphicx}
\graphicspath{{./images/}}

\usepackage{booktabs}
\usepackage{threeparttable}

\usepackage{hyperref}
\hypersetup{
    colorlinks   = true,
    linkcolor    = NavyBlue,
    citecolor    = BrickRed,
    urlcolor     = NavyBlue,
   pdftitle     = {Photon Propagation through Magnetar-Hosted Axion Clouds: Time Delays and Polarimetric Constraints},
   pdfauthor    = {M. M. Chaichian, B. A. Couto e Silva, B. L. S\'anchez-Vega}
}

\begin{document}

\title{Photon Propagation through Axion Clouds around Magnetized Compact Objects: Time Delays and Polarimetric Signatures}

\author{M. M. Chaichian}
\email{masud.chaichian@helsinki.fi}
\affiliation{Department of Physics and Helsinki Institute of Physics, University of Helsinki, P.O. Box 64, 00014 Helsinki, Finland.}

\author{B. A. Couto e Silva}
\email{brunoaces@ufmg.br}
\affiliation{Departamento de F\'isica, UFMG, Belo Horizonte, MG 31270-901, Brazil.}

\author{B. L. S\'anchez-Vega}
\email{bruce@fisica.ufmg.br}
\affiliation{Departamento de F\'isica, UFMG, Belo Horizonte, MG 31270-901, Brazil.}

\begin{abstract}
Temporal offsets between Gamma-Ray Bursts (GRBs) and high-energy neutrinos probe propagation effects in extreme astrophysical environments. We investigate whether such offsets can be generated by photon propagation through dense axion clouds gravitationally bound to strongly magnetized compact objects, such as canonical pulsars. Working within the Euler--Heisenberg effective theory extended by the axion sector, we derive the photon dispersion relations in a strong magnetic background permeated by an oscillating axion field. The magnetized vacuum is birefringent already at the Euler--Heisenberg level and the axion cloud superimposes density-dependent, time-dependent, and parity-odd structure on this baseline. The resulting geometry-dependent deviations from luminal propagation yield kinematic time delays reaching $\Delta t_{\perp} \simeq 9.3 \times 10^{-12}\,\mathrm{s}$ for $\mathbf{k} \perp \mathbf{B}$, far too small to account for macroscopic multimessenger offsets, so that propagation through such environments cannot, by itself, rule out Lorentz-invariance violation as an explanation for GRB offsets. In the polarization sector, we show that the axion-induced circular birefringence is an endpoint effect and therefore does not produce a leading net rotation for a complete vacuum-to-vacuum transit through a localized cloud. For configurations with a nonzero axion-field endpoint contrast, the same parity-odd phase defines a conditional environmental sensitivity benchmark. We find that this mechanism is optimally sensitive in the ultralight regime, yielding a benchmark reach of $g_{a\gamma\gamma} \lesssim 2.5 \times 10^{-10}\,\mathrm{GeV}^{-1}$ for canonical pulsar fields and axion masses near $m_a \sim 10^{-9}\,\mathrm{eV}$, and we identify the parity-odd (chiral) birefringence induced by the oscillating cloud as the signature that distinguishes the axion contribution from the calculable QED baseline.
\end{abstract}

\maketitle
\newpage

\tableofcontents


\newpage

\section{Introduction}

The astrophysical origin of high-energy neutrinos remains one of the central open problems in modern particle astrophysics. Gamma-Ray Bursts (GRBs) are among the leading candidate sources of high-energy neutrino emission, which is commonly expected to arise from photohadronic ($p\gamma$) interactions in relativistic fireballs \cite{Fireball1,Fireball2,Berezinsky1969,Stecker1978,Waxman1996,MiraldaEscude1996,Murase2006,Bustamante2015}. This possibility has received renewed attention following the recent KM3NeT report of the most energetic neutrino detected so far \cite{KM3NeT}. Even so, a statistically robust observational association between GRB activity and high-energy neutrino emission is still lacking. Large-scale detectors such as IceCube \cite{ICE2016,Aartsen2017,ICE2022} and ANTARES \cite{Antaresmuon} have not established significant correlations between gamma-ray emission and neutrino arrival times, and present data suggest that many reported temporal coincidences may be accidental \cite{JCAP2016,ANTICE}. In this context, any offset between electromagnetic and neutrino signals becomes a useful observable, since it can encode both intrinsic source physics and propagation effects along the line of sight.

One possible interpretation of such temporal offsets invokes Lorentz-invariance violation at extreme energies \cite{RodriguezMartinez2006,IceCubeLIV,Masudepjc,Cao2024,Song2024,Masudplb}. In these scenarios, often motivated by quantum-gravity constructions \cite{AmelinoCamelia} or string-inspired spacetime-foam models \cite{Mavromatos20105409}, the vacuum behaves effectively as a dispersive medium, inducing an energy dependence in the photon velocity. Before attributing any observed delay to the breakdown of a fundamental symmetry, however, it is necessary to quantify the propagation effects that arise within conventional astrophysical environments.

This standard-physics baseline has recently been examined for several diffuse media, including electron-positron plasmas, the Cosmic Microwave Background (CMB), and axion dark matter \cite{Masudepjc}. Although dispersion is then unavoidable in principle, the resulting delays for high-energy transients are extremely small, typically in the range $\mathcal{O}(10^{-23})$ to $\mathcal{O}(10^{-50})~\mathrm{s}$ depending on the background. Diffuse astrophysical media therefore do not provide delays of a size comparable to those discussed in present multimessenger candidates. This naturally motivates the study of localized high-density environments in which axion-photon interactions may be substantially enhanced.

A particularly interesting setting is provided by strongly magnetized compact objects, especially pulsars \cite{Adler1971,Tsai1974,Karbstein2013}. Recent work has shown that nonstationary pair-plasma discharges in the polar-cap region can efficiently source axions and that, for masses in the range $10^{-9}\,\mathrm{eV}\le m_a \le 10^{-4}\,\mathrm{eV}$, a significant fraction of the produced population may remain gravitationally bound to the star, gradually accumulating into dense axion clouds over astrophysical timescales \cite{Noordhuis2024}. Such clouds may reach local densities above $\mathcal{O}(10^{22})~\mathrm{GeV/cm^3}$, thereby defining a highly nontrivial environment for photon propagation within the magnetosphere.

In this setting, photon transport is governed by the combined presence of an external magnetic field and an oscillating axion background. While magnetospheric plasma may also affect the refractive properties of the medium, in the present work we isolate the contribution associated with QED vacuum polarization and axion-photon mixing. For the benchmark configuration adopted here, the external magnetic field remains below the critical QED scale,
\begin{equation}
B_{\mathrm{critical}}=\frac{m_e^2}{e}\simeq 4.4\times10^{13}\,\mathrm{G},
\label{eq:Bc}
\end{equation}
so that the Euler--Heisenberg description remains formally controlled as a weak-field expansion [\citenum{Karbstein2013}, \citenum{Masudplb}]. Within this regime, the magnetized vacuum behaves as an anisotropic birefringent medium, while the axion sector further modifies the photon dispersion relation, opening the possibility of geometry-dependent deviations from luminal propagation and polarization-dependent phase accumulation \cite{Optical}. Our analysis should therefore be viewed as an effective phenomenological description of the leading dispersive and birefringent structure of the propagation problem.

The aim of this work is twofold. First, we ask whether photon propagation through a pulsar-hosted axion cloud can generate GRB--neutrino time delays of observational relevance. Second, we determine what complementary birefringent constraints can be extracted from the same environment. To this end, we derive the modified photon dispersion relations and the corresponding group velocities in the effective magnetized axion background. We show that the local contribution to the photon delay can be enhanced up to $\mathcal{O}(10^{-12})~\mathrm{s}$, many orders of magnitude above the values expected in diffuse astrophysical backgrounds, yet still far below the $\mathcal{O}(1)~\mathrm{s}$ scale discussed in current multimessenger candidates. Our analysis therefore shows that this propagation mechanism is insufficient, by itself, to account for macroscopic GRB--neutrino offsets. At the same time, because the same anisotropic medium contains a parity-odd circular-birefringence channel, we use endpoint configurations with nonzero axion-field contrast to define an environmental sensitivity benchmark on the axion-photon coupling $g_{a\gamma\gamma}$.

The paper is organized as follows. In Sec.~\ref{theory}, we present the theoretical framework and derive the modified equations of motion and dispersion relations from the Euler--Heisenberg action extended by the axion sector. In Sec.~\ref{timedelay}, we compute the kinematic time delay for the canonical propagation modes. In Sec.~\ref{sec:polarization}, we analyze the birefringent properties of the medium and the resulting projected sensitivity in the axion parameter space. Finally, our conclusions are given in Sec.~\ref{sec:conclusions}.

\section{Theoretical Framework}
\label{theory}

In this section we formulate the effective description of photon propagation through a magnetized compact object environment permeated by a dense axion cloud. We begin by deriving the linearized field equations for electromagnetic and axionic fluctuations around prescribed background fields. We then construct the homogeneous propagation operator and use it to determine the local dispersion branches in the two canonical configurations, $\mathbf{k}\parallel\mathbf{B}$ and $\mathbf{k}\perp\mathbf{B}$. Our analysis is carried out within a local WKB/adiabatic approximation, in which the background fields vary slowly on the scale of the microscopic wavelength of the photon probe. We adopt Heaviside--Lorentz units throughout, setting $\hbar=c=1$. For notational clarity, $\mathbf{k}$ denotes the spatial wave vector, and $\mathbf{k}^2 \equiv \mathbf{k}\cdot\mathbf{k}$ its Euclidean norm squared.

\subsection{Theoretical Setup and Field Equations}
\label{sec:setup}

We study the propagation of a photon probe through the magnetosphere of a magnetized compact object hosting a dense axion cloud, with the aim of isolating the contributions of QED vacuum polarization \cite{Schwinger} and axion-photon mixing to the local dispersive and birefringent response of the medium.

For the low-energy photon modes considered here, with $\omega \ll m_e$, the coupled photon--axion system is described by the Euler--Heisenberg effective action \cite{EH1936} extended by the axion sector,
\begin{equation}
\label{eq:Leff}
\begin{aligned}
  \mathcal{L}_{\rm eff}&=
  \underbrace{-\tfrac14F_{\mu\nu}F^{\mu\nu}
  +\frac{\alpha^2}{90m_e^4}\Bigl[(F_{\mu\nu}F^{\mu\nu})^2+\tfrac74(F_{\mu\nu}\tilde F^{\mu\nu})^2\Bigr]}_{\mathcal{L}_{EH}}\\
  &+\tfrac12\partial_\mu \phi\,\partial^\mu \phi-\tfrac12m_a^2\phi^2
  -\underbrace{\frac{g_{a \gamma\gamma}}{4}\,\phi\,F_{\mu\nu}\tilde F^{\mu\nu}}_{\mathcal{L}_{a\gamma \gamma}}-J_\mu A^\mu.
\end{aligned}
\end{equation}
Here $\tilde F^{\mu\nu}=\tfrac12\epsilon^{\mu\nu\rho\sigma}F_{\rho\sigma}$ is the dual electromagnetic tensor, and $\phi$ denotes the axion field, which couples to the electromagnetic sector through the anomalous interaction term. 

The term $\mathcal{L}_{EH}$ encodes the one-loop QED vacuum-polarization correction induced by the background electromagnetic field, while $\mathcal{L}_{a\gamma\gamma}$ describes axion--photon mixing in the presence of macroscopic fields. Throughout this section we retain only the leading Euler--Heisenberg operators. Since our benchmark magnetic field is $B = 10^{12}\,\mathrm{G}$ \cite{Baring1998Radio-Quiet}, well below the critical QED scale in eq.\eqref{eq:Bc}, the weak-field truncation remains formally controlled. The coefficients appearing below therefore correspond to the standard one-loop weak-field Euler--Heisenberg response and are used here to capture the leading anisotropic, birefringent, and axion-induced structure of the local propagation problem.

To derive the modified Maxwell equations and the corresponding dispersion relations, we follow the field-decomposition strategy developed in refs.\cite{Helayel1,Helayel2} for axion--photon mixing in non-linear electrodynamics, adapting it to a localized and time-dependent magnetized compact object environment.

We decompose the fields into prescribed background configurations plus small probe fluctuations,
\begin{equation}
F^{\mu \nu}=F^{\mu \nu}_B+f^{\mu \nu},\qquad 
\phi=\phi_B+a,\qquad 
A^\mu=A^\mu_B+A^\mu_f.
\end{equation}
Here $f_{\mu\nu}=\partial_\mu A_{f\nu}-\partial_\nu A_{f\mu}$ denotes the propagating electromagnetic fluctuation around the prescribed background $F_B^{\mu\nu}$, while $a$ is the axion fluctuation around the background field $\phi_B$. The background fields are treated as prescribed classical configurations defining the local medium seen by the probe, rather than as a fully backreacted solution of the complete nonlinear system. We further assume that $(f_{\mu\nu},a)$ remain perturbatively small. 

For the purpose of determining the local propagation eigenmodes, we retain
the part of the second variation of the action that is bilinear in the
fluctuations $(f_{\mu\nu},a)$. Background-only terms and terms linear in
the fluctuations are not included in this propagation action. For
prescribed backgrounds that do not necessarily satisfy the fully
backreacted equations of motion, the omitted linear terms describe
externally driven contributions to the fluctuation fields. They do not,
however, modify the quadratic fluctuation operator, its polarization
tensor, or the homogeneous dispersion relations studied below.

The resulting bilinear propagation Lagrangian is
\begin{equation}
\begin{aligned}
\mathcal{L}_{\rm prop}^{(2)}
={}&
-\frac{1}{4}c_1 f_{\mu\nu}f^{\mu\nu}
-\frac{1}{4}
\bigl(c_2+g\phi_B(\mathbf{x},t)\bigr)
f_{\mu\nu}\widetilde f^{\mu\nu}
+\frac{1}{8}
Q_{B\mu\nu\kappa\lambda}
f^{\mu\nu}f^{\kappa\lambda}
\\
&+
\frac{1}{2}\partial_\mu a\,\partial^\mu a
-\frac{1}{2}m_a^2a^2
-\frac{g_{a \gamma\gamma}}{2}
a\widetilde F_B^{\mu\nu}f_{\mu\nu}.
\label{L2gen}
\end{aligned}
\end{equation}

The tensor $Q_{B\mu\nu\kappa\lambda}$ encodes the quadratic
electromagnetic response of the background,
\begin{equation}
\begin{aligned}
Q_{B\mu\nu\kappa\lambda}
={}&
d_1 F_{B\mu\nu}F_{B\kappa\lambda}
+d_2\widetilde F_{B\mu\nu}\widetilde F_{B\kappa\lambda} + d_3 \left(F_{B\mu\nu}\widetilde F_{B\kappa\lambda} + \widetilde F_{B\mu\nu}F_{B\kappa\lambda}\right),
\end{aligned}
\end{equation}
where the coefficients are obtained from derivatives of the
Euler--Heisenberg Lagrangian with respect to the electromagnetic
invariants
\begin{equation}
\mathcal{F}
=
-\frac14 F_{\mu\nu}F^{\mu\nu},
\qquad
\mathcal{G}
=
-\frac14 F_{\mu\nu}\widetilde F^{\mu\nu}.
\end{equation}
With these conventions, the Euler--Heisenberg
Lagrangian can be written as
\begin{equation}
\mathcal{L}_{EH}
=
\mathcal{F}
+
16\epsilon\,\mathcal{F}^{2}
+
28\epsilon\,\mathcal{G}^{2},
\qquad
\epsilon\equiv\frac{\alpha^{2}}{90m_{e}^{4}}.
\label{eq:LEH_invariants}
\end{equation}
The response coefficients are therefore defined by
\begin{align}
c_1
&=
\left.
\frac{\partial\mathcal{L}_{EH}}{\partial\mathcal{F}}
\right|_{B},
&
c_2
&=
\left.
\frac{\partial\mathcal{L}_{EH}}{\partial\mathcal{G}}
\right|_{B},
\\
d_1
&=
\left.
\frac{\partial^{2}\mathcal{L}_{EH}}
{\partial\mathcal{F}^{2}}
\right|_{B},
&
d_2
&=
\left.
\frac{\partial^{2}\mathcal{L}_{EH}}
{\partial\mathcal{G}^{2}}
\right|_{B},
&
d_3
&=
\left.
\frac{\partial^{2}\mathcal{L}_{EH}}
{\partial\mathcal{F}\partial\mathcal{G}}
\right|_{B}.
\end{align}

A final ingredient entering these coefficients is the modeling of the
axion cloud itself. In the standard dark-matter context, an axion
condensate with large occupation number is well described as a coherent
classical field oscillating at frequency $\omega\simeq m_a$. Although the
axion population considered here is of astrophysical rather than
primordial origin, the same classical-field description applies once the
gravitationally bound cloud reaches sufficiently large occupation number. We therefore parametrize the local cloud background
as
\begin{equation}
\phi_B(\mathbf{x},t)
=
\phi_0(\mathbf{x})\cos(m_at),
\label{axion_field}
\end{equation}
with amplitude fixed by the local energy density,
\begin{equation}
\rho_a(\mathbf{x})
\simeq
\frac{1}{2}m_a^2\phi_0^2(\mathbf{x}).
\end{equation}
In practice, $\phi_0(\mathbf{x})$ may vary over magnetospheric scales,
but within the local mode analysis it enters parametrically through its
value at the propagation point. The crucial difference with respect to
the diffuse Galactic halo is the density scale: whereas the local halo
density is typically
$\rho_{\rm DM}\sim0.3$--$0.4~\mathrm{GeV/cm^3}$
\cite{Catena2009,Iocco2011,Sofue2020}, the magnetospheric cloud
considered here may reach
$\rho_a\sim10^{22}~\mathrm{GeV/cm^3}$ \cite{Noordhuis2024}.

It is also important to distinguish the electric fields responsible for
axion production from those relevant to the optical properties of the
established cloud. As discussed in ref.\cite{Noordhuis2024}, the cloud
is initially sourced by nonstationary gap fields, $E_{\rm gap}$, in the
polar-cap region. Once the cloud has formed, however, the coherent
oscillation of $\phi_B(t)$ itself modifies the effective electromagnetic
response of the medium. In the presence of the strong static magnetic
field $\mathbf{B}$, the oscillating axion background acts as an effective
current source through the modified Amp\`ere law,
\begin{equation}
\mathbf{J}_{\rm eff}
=
-g_{a\gamma\gamma}\dot{\phi}_B\mathbf{B}.
\end{equation}
Self-consistency of Maxwell's equations then requires a compensating
displacement current, which induces an oscillatory electric field aligned
with the magnetic field lines. At the level of the local coherent-field
approximation, and up to an irrelevant phase convention, this induced
electric field may be estimated as
\begin{equation}
\mathbf{E}_\parallel(t)
=
-\frac{
g_{a\gamma\gamma}B\sqrt{2\rho_a}
}{
m_a
}
\cos(m_at)\,\hat{\mathbf{B}},
\label{E_induced}
\end{equation}
where $\rho_a$ is the local cloud density and $\hat{\mathbf{B}}$ is the
unit vector along the magnetic-field direction. In what follows,
$\mathbf{B}$ denotes the prescribed pulsar magnetic
field, while $\mathbf{E}$ denotes this induced local electric component
entering the electromagnetic invariants. This field is distinct from the
original gap field that sourced the axions, but it is the relevant
electric background for the local optical response analyzed here, since
it renders the invariant
$\mathcal{G}\propto\mathbf{E}\cdot\mathbf{B}$
non-vanishing and time dependent.

For the standard Euler--Heisenberg Lagrangian, the response coefficients
become
\begin{equation}
\begin{aligned}
c_1
&=
1+16\epsilon
\left(
\vec{E}^{\,2}-\vec{B}^{\,2}
\right),
\qquad
c_2
=
56\epsilon
\left(
\vec{E}\cdot\vec{B}
\right),
\\
d_1
&=
32\epsilon,
\qquad
d_2
=
56\epsilon,
\qquad
d_3
=
0.
\end{aligned}
\end{equation}

The vanishing of $d_3$ reflects the parity-even structure of standard
QED. For the benchmark configuration adopted here, these coefficients
are the standard weak-field Euler--Heisenberg coefficients entering the
present truncation.

The dynamics of the coupled photon--axion fluctuations follow from the
stationary-action principle. Varying eq.~\eqref{L2gen} with respect to $A_f^\mu$ and $a$ yields the
homogeneous linearized equations governing probe propagation on top of the prescribed magnetized axion background.

Variation with respect to the gauge field gives
\begin{equation}
\partial^\mu
\left[
c_1f_{\mu\nu}
+
\bigl(c_2+g_{a \gamma\gamma}\phi_B(\mathbf{x},t)\bigr)
\widetilde f_{\mu\nu}
-
\frac{1}{2}
Q_{B\mu\nu\kappa\lambda}f^{\kappa\lambda}
+
g_{a \gamma\gamma}a\widetilde F_{B\mu\nu}
\right]
=
0,
\label{eqmovA}
\end{equation}
whereas variation with respect to the axion fluctuation yields
\begin{equation}
(\Box+m_a^2)a
=
g_{a \gamma\gamma}(\mathbf{e}\cdot\mathbf{B})
+
g_{a \gamma\gamma}(\mathbf{b}\cdot\mathbf{E}).
\label{eqmov_axion}
\end{equation}
Here $\mathbf{E}$ and $\mathbf{B}$ denote the prescribed background
fields, whereas $\mathbf{e}$ and $\mathbf{b}$ denote the probe
fluctuations. Both equations are homogeneous in the fluctuation fields.
The spacetime dependence of the prescribed backgrounds nevertheless
enters the propagation operator through derivatives of the corresponding
background-dependent coefficients.

Combining eq.~\eqref{eqmovA} with the Bianchi identity, the system can
be written in Maxwell form as
\begin{equation}
\begin{aligned}
\nabla\cdot\mathbf{D}
&=
0,
\\
\nabla\cdot\mathbf{b}
&=
0,
\\
\nabla\times\mathbf{e}
+
\frac{\partial\mathbf{b}}{\partial t}
&=
0,
\\
\nabla\times\mathbf{H}
-
\frac{\partial\mathbf{D}}{\partial t}
&=
0,
\end{aligned}
\label{MMEq}
\end{equation}
where the medium response is encoded in the constitutive relations for
$\mathbf{D}$ and $\mathbf{H}$. These receive contributions from both QED
vacuum polarization, through the coefficients $c_i$ and $d_i$, and the
axion background:
\begin{equation}
\begin{aligned}
\mathbf{D}
={}&
c_1\mathbf{e}
+
\bigl(c_2+g_{a \gamma\gamma}\phi_B(\mathbf{x},t)\bigr)\mathbf{b}
+
d_1\mathbf{E}
\left(
\mathbf{E}\cdot\mathbf{e}
\right)
+
d_2\mathbf{B}
\left(
\mathbf{B}\cdot\mathbf{e}
\right)\\
&-d_1\mathbf{E}
\left(
\mathbf{B}\cdot\mathbf{b}
\right)
+
d_2\mathbf{B}
\left(
\mathbf{E}\cdot\mathbf{b}
\right)
+
g_{a \gamma\gamma}a\mathbf{B},
\end{aligned}
\label{eq:D_constitutive}
\end{equation}
\begin{equation}
\begin{aligned}
\mathbf{H}
={}&
c_1\mathbf{b}
-
\bigl(c_2+g_{a \gamma\gamma}\phi_B(\mathbf{x},t)\bigr)\mathbf{e}
-
d_1\mathbf{B}
\left(
\mathbf{B}\cdot\mathbf{b}
\right)
-
d_2\mathbf{E}
\left(
\mathbf{E}\cdot\mathbf{b}
\right)\\
&+d_1\mathbf{B}
\left(
\mathbf{E}\cdot\mathbf{e}
\right)
-
d_2\mathbf{E}
\left(
\mathbf{B}\cdot\mathbf{e}
\right)
-
g_{a \gamma\gamma}a\mathbf{E}.
\end{aligned}
\label{eq:H_constitutive}
\end{equation}

It is important to emphasize that the background electric field entering
these expressions is not an independent external input, but is entirely
induced by the oscillating axion cloud, as defined in
eq.~\eqref{E_induced}. Accordingly, in the limit where the axion sector
is switched off, $g_{a \gamma\gamma}\to0$, the induced electric background vanishes
consistently, $\mathbf{E}\to0$. In this limit, the explicit
axion-dependent terms disappear, and the constitutive relations reduce
smoothly to those of the standard Euler--Heisenberg vacuum in a purely
magnetic background.

The spacetime dependence of $\phi_B$ does not introduce an external
source into the propagation problem defined by
eq.~\eqref{L2gen}. Instead, derivatives of $\phi_B$ multiply the probe
fields and therefore enter the homogeneous fluctuation operator. To
extract the local propagation modes, we now adopt a WKB description in
which
\begin{equation}
\lambda_{\rm probe}
\ll
L_\phi,\;L_B,
\qquad
|\nabla\phi_0|
\ll
m_a|\phi_0|,
\end{equation}
where $L_\phi$ and $L_B$ denote the characteristic spatial variation
scales of the axion background and the magnetic field, respectively.
The background fields may therefore be treated as locally constant,
while the explicit temporal modulation of $\phi_B$ is retained.

Neglecting the subleading spatial-gradient terms proportional to
$\nabla\phi_B$, the Fourier amplitudes satisfy
\begin{equation}
\begin{alignedat}{2}
\mathbf{k}\cdot\mathbf{D}_0
&=
0,
&\qquad
\mathbf{k}\times\mathbf{e}_0
&=
\omega\mathbf{b}_0,
\\
\mathbf{k}\cdot\mathbf{b}_0
&=
0,
&\qquad
\mathbf{k}\times\mathbf{H}_0
+
\omega\mathbf{D}_0
&=
-ig_{a \gamma\gamma}\dot{\phi}_B\mathbf{b}_0,
\end{alignedat}
\label{eq:fourier_maxwell}
\end{equation}
together with
\begin{equation}
\left(
\mathbf{k}^2-\omega^2+m_a^2
\right)a_0
=
g_{a \gamma\gamma}\mathbf{B}\cdot\mathbf{e}_0
+
g_{a \gamma\gamma}\mathbf{E}\cdot\mathbf{b}_0.
\label{eq:fourier_system}
\end{equation}
Here $D_{0i}$ and $H_{0i}$ denote the Fourier amplitudes of
$\mathbf{D}$ and $\mathbf{H}$. A closely related system of equations was
derived in ref.~\cite{Ouellet2019}. The term proportional to
$\dot{\phi}_B\mathbf{b}_0$ is linear in the probe fluctuation and is
therefore part of the homogeneous propagation operator rather than an
external source.

A key step is to eliminate the axion fluctuation $a_0$ in favor of the
electromagnetic amplitudes. At the linearized level, this amounts to
integrating out the \emph{fluctuating} axion mode, not the background
field itself. Solving the Klein--Gordon equation for $a_0$ and
substituting the result back into the constitutive relations allows us to
express the response of the medium purely in terms of $e_{0j}$ and
$b_{0j}$,
\begin{align}
D_{0i}(\mathbf{k},\omega)
&=
\epsilon_{ij}(\mathbf{k},\omega)e_{0j}
+
\sigma_{ij}(\mathbf{k},\omega)b_{0j},
\\
H_{0i}(\mathbf{k},\omega)
&=
-\sigma_{ji}(\mathbf{k},\omega)e_{0j}
+
(\mu^{-1})_{ij}(\mathbf{k},\omega)b_{0j}.
\end{align}

The effective medium is thus encoded in the susceptibility tensors. The
electric permittivity tensor
$\epsilon_{ij}(\mathbf{k},\omega)$ and the magnetoelectric tensor
$\sigma_{ij}(\mathbf{k},\omega)$ contain the resonant axion contribution
through the pole at
$\mathbf{k}^2-\omega^2+m_a^2=0$. Their explicit components are
\begin{align}
\epsilon_{ij}(\mathbf{k},\omega)
&=
c_1\delta_{ij}
+
d_1E_iE_j
+
d_2B_iB_j
+
\frac{
g_{a \gamma\gamma}^2B_iB_j
}{
\mathbf{k}^2-\omega^2+m_a^2
},
\\
\sigma_{ij}(\mathbf{k},\omega)
&=
-\bigl(c_2+g\phi_B\bigr)\delta_{ij}
-
d_1E_iB_j
+
d_2B_iE_j
+
\frac{
g_{a \gamma\gamma}^2B_iE_j
}{
\mathbf{k}^2-\omega^2+m_a^2
}.
\end{align}
Similarly, the inverse magnetic permeability tensor
$(\mu^{-1})_{ij}$ describes the magnetic response of the vacuum
modified by the axion interaction:
\begin{equation}
(\mu^{-1})_{ij}
=
c_1\delta_{ij}
-
d_1B_iB_j
-
d_2E_iE_j
-
\frac{
g_{a \gamma\gamma}^2E_iE_j
}{
\mathbf{k}^2-\omega^2+m_a^2
}.
\end{equation}

To determine the propagation eigenmodes, we eliminate the magnetic
fluctuation $\mathbf{b}_0$ using Faraday's law,
\begin{equation}
\mathbf{b}_0
=
\frac{1}{\omega}
\left(
\mathbf{k}\times\mathbf{e}_0
\right),
\end{equation}
and substitute this relation into the constitutive equations. Inserting
the resulting expressions for
$\mathbf{D}_0(\mathbf{e}_0)$ and
$\mathbf{H}_0(\mathbf{e}_0)$ into the Amp\`ere--Maxwell equation yields
a homogeneous wave equation for the electric amplitude,
\begin{equation}
M^{ij}{e_0}_j
=
0,
\label{eq:homogeneous_wave_equation}
\end{equation}
with
\begin{equation}
\begin{aligned}
M^{ij}
={}&
a\delta^{ij}
+
bk^ik^j
+
c_BB^iB^j
+
c_EE^iE^j
-
\frac{ig_{a \gamma\gamma}}{c_1}
\dot{\phi}_B
\epsilon^{ijm}k_m
\\
&+
d_B
(\mathbf{B}\cdot\mathbf{k})
\left(
B^ik^j+B^jk^i
\right)
+
d_E
(\mathbf{E}\cdot\mathbf{k})
\left(
E^ik^j+E^jk^i
\right)
\\
&-
d_B\omega
\left[
E^i(\mathbf{B}\times\mathbf{k})^j
+
E^j(\mathbf{B}\times\mathbf{k})^i
\right]
\\
&+
d_E\omega
\left[
B^i(\mathbf{E}\times\mathbf{k})^j
+
B^j(\mathbf{E}\times\mathbf{k})^i
\right],
\label{Mtensor}
\end{aligned}
\end{equation}
where
\begin{equation}
d_B
\equiv
\frac{d_1}{c_1},
\qquad
d_E
\equiv
\frac{d_2}{c_1}
+
\frac{
g_{a \gamma\gamma}^2
}{
c_1
\left(
\mathbf{k}^2-\omega^2+m_a^2
\right)
},
\end{equation}
\begin{equation}
a
\equiv
\omega^2-\mathbf{k}^2
+
d_B(\mathbf{k}\times\mathbf{B})^2
+
d_E(\mathbf{k}\times\mathbf{E})^2,
\end{equation}
\begin{equation}
b
\equiv
1-d_B\mathbf{B}^2-d_E\mathbf{E}^2,
\end{equation}
\begin{equation}
c_B
\equiv
d_E\omega^2-d_B\mathbf{k}^2,
\qquad
c_E
\equiv
d_B\omega^2-d_E\mathbf{k}^2.
\end{equation}

A useful structural feature of this result is that the isotropic part of
the magnetoelectric tensor, namely the term proportional to $c_2$,
cancels identically from the homogeneous propagation operator $M^{ij}$.
Thus, the spatially isotropic pseudoscalar contribution does not by
itself generate nontrivial mode splitting at the level of the
homogeneous wave equation; the anisotropic structure instead arises from
the external fields and from the explicitly time-dependent axion
background.

It is instructive to compare $M^{ij}$ with dispersion matrices obtained
in static axion backgrounds and in Carroll--Field--Jackiw-type
electrodynamics \cite{Carroll1990,Helayel2}. In the static-background case, such as in
ref.~\cite{Helayel1}, the dispersion matrix is symmetric. By contrast,
the time dependence of $\phi_B$ generates the antisymmetric term
proportional to $\epsilon^{ijm}k_m$. The analogy with Lorentz- and
CPT-violating electrodynamics is purely kinematical: no fundamental
Lorentz violation is assumed here, since the preferred direction arises
from a physical time-dependent background. This parity-odd structure
lifts the degeneracy between opposite circular polarizations and
underlies the birefringent effects discussed in
Sec.~\ref{sec:polarization}.

The source-driven response associated with terms linear in the fluctuations, which were excluded from the propagation action~\eqref{L2gen}, lies outside the scope of the present analysis. Instead, we focus strictly on the free propagation eigenmodes supported by the effective medium.

\subsection{Dispersion Relations and Canonical Propagation Modes}
\label{disprel}

To determine the propagation observables relevant for time-of-flight, in particular, the group velocity, we must extract the local dispersion relations of these eigenmodes. They follow directly from the homogeneous propagation equation~\eqref{eq:homogeneous_wave_equation}, which defines the free wave solutions of the local medium. Non-trivial propagation modes exist only when the determinant of the propagation operator vanishes,
\begin{equation}
\det(M^{ij})=0.
\label{detM}
\end{equation}
Solving this condition yields the local dispersion branches, from which the corresponding group velocities can be determined.

Because the axion background is explicitly time dependent, the resulting dispersion relations should be interpreted as local instantaneous branches within the WKB/adiabatic approximation introduced above. In practice, $\phi_B(t)$ and $\dot{\phi}_B(t)$ are treated as locally constant over the microscopic oscillation time of the photon probe, so that one can define instantaneous propagation modes. Since the axion fluctuation has been integrated out, the determinant may contain resonant denominators inherited from that algebraic elimination. These structures should not be identified automatically with observable photon branches; the physically relevant modes are those continuously connected to the vacuum photon sector in the weak-background limit.

We analyze the system in the two canonical geometries selected by the external magnetic field: propagation parallel to the field lines (longitudinal mode) and propagation perpendicular to them (transverse mode). These two limiting cases provide the clearest physical characterization of the effective medium. The longitudinal configuration is sufficiently simple to admit closed-form expressions for the local branches, whereas the transverse configuration leads to a higher-order characteristic equation that must be handled numerically.

\subsubsection{Longitudinal mode}

In the first canonical configuration, we consider propagation along the magnetic axis of the magnetized compact object. In this longitudinal geometry, the wave vector $\mathbf{k}$ is parallel to both the background magnetic field $\mathbf{B}$ and the induced electric field $\mathbf{E}_\parallel$, so that
\begin{equation}
\mathbf{k}=k_z\,\hat{\mathbf{z}},
\qquad
\mathbf{B}=B_z\,\hat{\mathbf{z}},
\qquad
\mathbf{E}=E_z\,\hat{\mathbf{z}}.
\end{equation}
Substituting these kinematical conditions into the propagation tensor of eq.\eqref{Mtensor}, the determinant factorizes into dynamically distinct sectors,
\begin{equation}
\label{detM0}
\begin{split}
\det(M_{\parallel})
={}&
\omega^2
\left[
c_1^2
\left(
k_z^2-\omega^2
\right)^2
-
g_{a \gamma\gamma}^2\dot{\phi}_B^{\,2}k_z^2
\right]
\\
&\times
\frac{
B_z^2
\left[
d_2
\left(
k_z^2+m_a^2-\omega^2
\right)
+
g_{a \gamma\gamma}^2
\right]
+
\left(
c_1+d_1E_z^2
\right)
\left(
k_z^2+m_a^2-\omega^2
\right)
}{
c_1^3
\left(
k_z^2+m_a^2-\omega^2
\right)
}
=0.
\end{split}
\end{equation}

Besides the trivial root $\omega=0$, corresponding to a constraint mode, the non-trivial branches are
\begin{equation}
\omega_1
=
\pm
\sqrt{
k_z^2
-
\frac{k_z}{c_1}
\dot{\phi}_B g_{a \gamma\gamma}
},
\end{equation}
\begin{equation}
\omega_2
=
\pm
\sqrt{
k_z^2
+
\frac{k_z}{c_1}
\dot{\phi}_B g_{a \gamma\gamma}
},
\end{equation}
\begin{equation}
\omega_3
=
\pm
\sqrt{
k_z^2+m_a^2
+
\frac{
B_z^2g_{a \gamma\gamma}^2
}{
B_z^2d_2+c_1+d_1E_z^2
}
}.
\end{equation}

The factorization of $\det(M_\parallel)$ reflects the presence of dynamically distinct sectors in the coupled photon--axion system. The branches $\omega_{1,2}$ are the photon-like modes of the longitudinal configuration: they remain continuously connected to the vacuum light cone in the limit of vanishing background-induced corrections and are therefore the relevant branches for the photon time-of-flight observable considered in this work. Their corrections scale linearly with $g_{a \gamma\gamma}\dot{\phi}_B$, but the resulting deformation of the light cone is parametrically tiny for the benchmark configuration adopted here.

The branch $\omega_3$ is qualitatively different. It belongs to the mixed photon--axion sector, depends explicitly on the axion mass, and remains gapped in the local analysis. It is therefore useful for characterizing the full spectrum of the coupled system, but it is not the primary branch governing the arrival-time observable of a photon signal continuously connected to the vacuum electromagnetic mode.

For the photon-like longitudinal branches $\omega_{1,2}$, the corresponding group velocities remain extremely close to the vacuum value. In particular, the local longitudinal photon sector differs from luminal propagation only at a parametrically tiny level, so that the cumulative time delay in the parallel configuration is negligible for the benchmark parameters considered here.

For numerical estimates, the values quoted below should be understood as benchmark results within the present effective truncation, rather than as precision predictions of strong-field QED. The quantities $E_z$ and $\dot{\phi}_B$ entering the local dispersion relations are evaluated at their respective peak amplitudes and therefore define an upper-envelope benchmark rather than values attained simultaneously at a single phase of the axion oscillation. The corresponding input parameters are summarized in Table~\ref{tab:benchmark}.

\begin{table}[htbp]
\centering
\renewcommand{\arraystretch}{1.9}
\begin{tabular}{lll}
\hline
\textbf{Quantity} & \textbf{Symbol} & \textbf{Value / Definition} \\
\hline
\multicolumn{3}{l}{\itshape Input Parameters} \\
Axion mass & $m_a$ & $\left(10^{-9}-10^{-4}\right) \mathrm{eV}$ \\
Axion coupling & $g_{a \gamma\gamma}$ & $10^{-11}\ \mathrm{GeV}^{-1}$ \\
Background magnetic field & $B_z$ & $10^{12}\ \mathrm{G}$ \\
Photon wave number & $k$ & $100\ \mathrm{keV}$ \\
\hline
\multicolumn{3}{l}{\itshape Axion Background} \\
Axion energy density & $\rho_a$ & $10^{22}\ \mathrm{GeV\,cm^{-3}}$ \\
Peak axion amplitude & $\phi_0$ & $\dfrac{\sqrt{2\rho_a}}{m_a}$ \\
Peak induced electric field & $E_z$ & $g_{a \gamma\gamma}B_z\phi_0$ \\
Peak axion time derivative & $|\dot{\phi}_B|$ & $m_a\phi_0=\sqrt{2\rho_a}$ \\
\hline
\multicolumn{3}{l}{\itshape Non-linear QED} \\
Euler--Heisenberg coefficient
& $\epsilon$
& $\simeq8.68\times10^{6}\ \mathrm{GeV}^{-4}$ \\
Effective electromagnetic coefficient
& $c_1$
& $1+16\epsilon(E_z^2-B_z^2)$ \\
Nonlinear-response coefficient
& $d_1$
& $32\epsilon$ \\
Nonlinear-response coefficient
& $d_2$
& $56\epsilon$ \\
\hline
\end{tabular}
\caption{Benchmark input parameters and derived quantities used in the axion-electrodynamics analysis. Natural units with $\hbar=c=1$ are assumed, and all quantities are converted to a common unit system before numerical evaluation. The quoted values of $E_z$ and $|\dot{\phi}_B|$ are peak amplitudes and do not correspond to the same phase of the axion oscillation.\label{tab:benchmark}}
\end{table}

For definiteness, we evaluate the group velocity on a representative positive-frequency photon-like branch, which remains continuously connected to the vacuum electromagnetic limit. The corresponding group velocity is given by
\begin{equation}
v_{g,\parallel}
=
\frac{
2c_1k_z+g_{a \gamma\gamma}\dot{\phi}_B
}{
2\sqrt{
c_1k_z
\left(
c_1k_z+g_{a \gamma\gamma}\dot{\phi}_B
\right)
}
},
\end{equation}
which can be expressed in the parametric form
\begin{equation}
\boxed{
v_{g,\parallel}
\simeq
\left(
1+9.06\times10^{-35}
\right)c.
}
\label{vpara}
\end{equation}
We note that this result is formally superluminal ($v_{g,\parallel}>c$), corresponding to a microscopic time advance rather than a physical time delay. In the present effective description, this behavior reflects the presence of a classical time-dependent axion background, as discussed in related CFJ-like formulations \cite{Helayel2}. It does not imply a violation of causality in the underlying theory, since the signal velocity remains bounded by $c$. At the same time, the associated birefringent correction in this longitudinal configuration remains parametrically negligible for astrophysical observables.

The benchmark evaluation confirms that the photon-like longitudinal branch remains effectively luminal. Kinematically, a photon traversing the axionic and magnetic medium of the magnetosphere in this parallel configuration therefore experiences a negligible cumulative arrival-time shift, as will be quantified in Sec.~\ref{timedelay}. This strong suppression is consistent with previous analyses of vacuum dispersion below the Schwinger scale, where deviations from luminal propagation remain extremely small in astrophysical settings \cite{Masudepjc,Masudplb}. To determine whether this quasi-luminal behavior is generic or instead a consequence of the longitudinal alignment, we now turn to the transverse configuration.

\subsubsection{Transverse mode}
\label{transversal}

We next consider propagation perpendicular to the external magnetic field,
with
\begin{equation}
\mathbf{B}=B_z\,\hat{\mathbf z},
\qquad
\mathbf{E}=E_z\,\hat{\mathbf z},
\qquad
\mathbf{k}=k_x\,\hat{\mathbf x}.
\end{equation}
In this geometry, $\mathbf{k}\cdot\mathbf{B}=0$, and substituting the
background fields into eq.~\eqref{Mtensor} gives
\begin{equation}
\begin{aligned}
\det(M_\perp)
={}&
\omega^2
\Bigg\{
\left[
\omega^2-k_x^2
+
\frac{k_x^2}{c_1}
\left(
d_1B_z^2
+
\left[
d_2+
\frac{g_{a \gamma\gamma}^2}
{k_x^2+m_a^2-\omega^2}
\right]
E_z^2
\right)
\right]
\\
&\times
\left[
\omega^2-k_x^2
+
\frac{\omega^2}{c_1}
\left(
\left[
d_2+
\frac{g_{a \gamma\gamma}^2}
{k_x^2+m_a^2-\omega^2}
\right]
B_z^2
+
d_1E_z^2
\right)
\right]
\\
&-
\frac{k_x^2}{c_1^2}
\left[
\omega^2B_z^2E_z^2
\left(
d_2-d_1+
\frac{g_{a \gamma\gamma}^2}
{k_x^2+m_a^2-\omega^2}
\right)^2+g_{a \gamma\gamma}^2\dot{\phi}_B^{\,2}
\right]
\Bigg\}
=0.
\label{DetMperp}
\end{aligned}
\end{equation}
The poles at
$k_x^2+m_a^2-\omega^2=0$ are inherited from the algebraic elimination of
the axion fluctuation. They are treated separately from the photon-like
zeros of the determinant.

Unlike the longitudinal determinant, eq.~\eqref{DetMperp} does not lead to
useful closed-form photon branches. We therefore determine the two
photon-like solutions numerically, following both their eigenvalues and
polarization eigenvectors as the background corrections are continuously
turned on. This procedure distinguishes the two polarization branches and
prevents the axion pole or auxiliary solutions from being misidentified as
photon modes. The resulting dispersions are understood as local
instantaneous relations within the WKB approximation.

For the branch continuously connected to the photon mode with $\mathbf e\perp\mathbf B$, the benchmark parameters of
Table~\ref{tab:benchmark} give
\begin{equation}
\boxed{
v_{g,T \perp}
\simeq
\left(
1-5.3\times10^{-8}
\right)c,
}
\label{vperpperp}
\end{equation}
and for the mode $\mathbf e\parallel\mathbf B$ give
\begin{equation}
\boxed{
v_{g,T \parallel}
\simeq
\left(
1-9.3\times10^{-8}
\right)c
.}
\label{vperppara}
\end{equation}
The physical origin of this correction follows directly from the
Euler--Heisenberg limit. Setting
\begin{equation}
\mathbf{E}=0,
\qquad
g_{a \gamma\gamma}=0,
\qquad
\mathbf{k}\perp\mathbf{B},
\end{equation}
the determinant factorizes into the two linear photon polarizations. To
leading order in the weak-field expansion, their refractive indices are
\begin{equation}
n_{\mathbf e\perp\mathbf B}-1
=
16\epsilon B_z^2
+
\mathcal{O}
\left(
(\epsilon B_z^2)^2
\right),
\label{eq:QED_index_perpendicular}
\end{equation}
and
\begin{equation}
n_{\mathbf e\parallel\mathbf B}-1
=
28\epsilon B_z^2
+
\mathcal{O}
\left(
(\epsilon B_z^2)^2
\right).
\label{eq:QED_index_parallel}
\end{equation}
Their difference is the standard Euler--Heisenberg vacuum-birefringence
splitting,
\begin{equation}
\Delta n_{\rm QED}
\equiv
n_{\mathbf e\parallel\mathbf B}
-
n_{\mathbf e\perp\mathbf B}
=
12\epsilon B_z^2
+
\mathcal{O}
\left(
(\epsilon B_z^2)^2
\right).
\label{eq:QED_birefringence_check}
\end{equation}

For $B_z=10^{12}\,\mathrm{G}$, these expressions give
\begin{equation}
n_{\mathbf e\perp\mathbf B}-1
\simeq
5.3\times10^{-8},
\qquad
n_{\mathbf e\parallel\mathbf B}-1
\simeq
9.3\times10^{-8},
\end{equation}
and hence
\begin{equation}
\Delta n_{\rm QED}
\simeq
4.0\times10^{-8}.
\end{equation}
The two numerical photon branches must approach these values when the axion
sector is switched off, providing a direct check of the root-tracking
procedure described in Appendix~\ref{numerics}.

The result in eq.~\eqref{vperpperp} is therefore controlled by the
Euler--Heisenberg correction to the
$\mathbf e\perp\mathbf B$ branch. Indeed, for $E_z\to0$, this branch obeys
\begin{equation}
\omega^2-k_x^2
+
\frac{d_1B_z^2}{c_1}k_x^2
=
0,
\end{equation}
which reproduces
$n_{\mathbf e\perp\mathbf B}-1\simeq16\epsilon B_z^2$.
The $\mathbf e\parallel\mathbf B$ branch instead contains the axion pole
through
\begin{equation}
d_2+
\frac{g_{a \gamma\gamma}^2}
{k_x^2+m_a^2-\omega^2},
\end{equation}
while the time-dependent background introduces the off-diagonal coupling
proportional to $g_{a \gamma\gamma}\dot{\phi}_B$.

The contrast between the longitudinal and transverse group velocities is
therefore driven primarily by the anisotropic Euler--Heisenberg response of
the magnetized vacuum. The axion sector adds resonant photon--axion mixing
to the $\mathbf e\parallel\mathbf B$ branch and a parity-odd coupling between
the transverse polarization sectors, while its effect on the
$\mathbf e\perp\mathbf B$ benchmark branch is subleading.

We now use these longitudinal and transverse group velocities to determine
the corresponding kinematic time delays.

\section{Calculating the Time Delay}
\label{timedelay}

To translate the modified dispersion relations of the previous section into an observable, we compute the arrival-time difference between a GRB photon and its associated neutrino. The aim is not to reconstruct a specific event, but to test, within a controlled benchmark, whether the propagation effects derived above can produce a measurable multimessenger delay.

A natural benchmark is the coincidence between trigger bn140807500 and a high-energy neutrino candidate reported in multimessenger searches \cite{Abbasi_2024}. Among the bursts in the GeV--TeV sample, this event showed the most significant combined temporal and angular correlation: the reconstructed track, of energy $E_\nu\simeq221\,\mathrm{GeV}$, arrived within a $100$-s window and only $2.3^\circ$ from the GRB localization. Its post-trial background probability, $p=0.097$, does not establish a physical association, but the event remains a useful reference point \cite{Catalog_2020}. No spectroscopic redshift is available; since its temporal structure and fluence match the short-GRB population, whose redshift distribution typically peaks near $z\sim0.5$ \cite{Berger2014}, we adopt this value as a population-based benchmark.

The configuration we analyze is a background GRB whose photon and neutrino both traverse an intervening magnetized compact-object environment. This is the setting in which the propagation delay is cleanly defined, because the two messengers share a common emission event that fixes a common $t=0$; were the photons instead produced inside the magnetosphere, the offset would be entangled with emission radii, opacity, and pair-cascade dynamics. The likelihood of such a line-of-sight alignment is examined separately in Appendix~\ref{sec:magnetar_model}.

In a spatially flat $\Lambda$CDM cosmology, the comoving distance to redshift $z$ is
\begin{equation}
D_C(z)
=\frac{c}{H_0}\int_0^z
\frac{dz'}{\sqrt{\Omega_m(1+z')^3+\Omega_\Lambda}},
\end{equation}
and with the Planck 2018 parameters \cite{Planck2018}, $H_0=67.4~\mathrm{km\,s^{-1}\,Mpc^{-1}}$, $\Omega_m=0.315$, $\Omega_\Lambda=0.685$, this gives $D_C(0.5)\simeq1.95~\mathrm{Gpc}$, which sets the propagation baseline. We will treat the neutrino kinematic term in flat space since a fully consistent FRW treatment leaves the total delay unchanged at the quoted precision, as expected, because the neutrino mass correction is parametrically tiny for $m_\nu\ll E_\nu$.

Two energies enter the estimate. The photon energy must respect the effective-theory regime of Sec.~\ref{sec:setup}, $\omega\ll m_e\simeq 511\,\mathrm{keV}$; since the prompt Fermi-GBM emission lies mostly in the $\mathcal{O}(10\text{--}300)\,\mathrm{keV}$ band,\footnote{The burst information was accessed through the Xamin interface of the HEASARC archive.} we take $E_\gamma=100\,\mathrm{keV}$. For the neutrino, $E_\nu\simeq221\,\mathrm{GeV}$ \cite{Abbasi_2024}, its velocity is
\begin{equation}
\beta_\nu\equiv\frac{v_\nu}{c}
\simeq 1-\underbrace{\frac{m_\nu^2}{2E_\nu^2}}_{d_\nu},
\label{neutrinovel}
\end{equation}
and adopting $m_\nu=0.05~\mathrm{eV}$ \cite{DESI2024,Planck2018} gives $d_\nu\simeq2.5\times10^{-26}$, so the neutrino is effectively luminal for the present purpose.

Finally, the trajectory separates into the effectively empty cosmological path and the local magnetized region the photon crosses. We model the latter as
\begin{equation}
D_{\rm int}=2\,r_{\rm psr},\qquad r_{\rm psr}=15~\mathrm{km},
\end{equation}
yielding $D_{\rm int}=30~\mathrm{km}$. This choice represents a conservative upper bound for the radius of a canonical pulsar \cite{Lattimer2001}, defining the innermost region over which the benchmark field and local dispersion relation apply, and it is distinct from the larger geometric radius used in Appendix~\ref{sec:magnetar_model} to estimate encounter probabilities. The remaining cosmological path length is therefore
\begin{equation}
D_{\rm empty}=D_C(z)-D_{\rm int}.
\end{equation}

With this decomposition, the total relative delay $\Delta t^{\rm tot} = t_\gamma - t_\nu$ splits naturally into a piece accumulated over this empty baseline and a piece accumulated inside the magnetized region,
\begin{equation}
\Delta t^{\rm tot}=
\underbrace{\frac{D_{\rm empty}}{c}\left(1-\beta_\nu^{-1}\right)}_{\text{vacuum term}}
+
\underbrace{\frac{D_{\rm int}}{c}\left(\beta_g^{-1}-\beta_\nu^{-1}\right)}_{\text{local term}},
\label{totaldelay}
\end{equation}
where $\beta_g=v_g/c$ denotes the group velocity on the relevant photon-like branch. Equation~\eqref{totaldelay} shows that the net temporal separation is determined by the competition between the accumulated kinematic delay of the massive neutrino over cosmological distances and the modified local dispersion of the photon inside the magnetized compact-object environment.

Evaluating eq.~\eqref{totaldelay} with our benchmark parameters reveals three distinct propagation scenarios, depending on the orientation and polarization of the photon relative to the external magnetic field. For the longitudinal ($\mathbf{k}\parallel\mathbf{B}$) case we have that
\begin{equation}
    \Delta t_{L\parallel}^{\rm tot}\simeq -5.021\times10^{-9}~\mathrm{s},
    \label{paradelays}
\end{equation}
and the two birefringent transverse ($\mathbf{k}\perp\mathbf{B}$) branches, are
\begin{equation}
\Delta t_{T_\parallel}^{\rm tot}\simeq -5.012\times10^{-9}~\mathrm{s},
\quad
\Delta t_{T\perp}^{\rm tot}\simeq -5.016\times10^{-9}~\mathrm{s}.
\label{eq:perp_delays}
\end{equation}
In every case, the negative sign indicates that the photon arrives slightly ahead of the neutrino. At these macroscopic scales, the kinematic neutrino correction accumulated over the full cosmological baseline ($\sim 1.95~\mathrm{Gpc}$) dominates the total observable delays, rendering them numerically very close. The slight spread among the three values is entirely the imprint of the local photon dispersion.

To isolate the effect of the medium, it is useful to extract the delay accumulated exclusively across the interaction region. The local contribution is
\begin{equation}
\Delta t^{\rm int}=
\frac{D_{\rm int}}{c}\left(\beta_g^{-1}-\beta_\nu^{-1}\right).
\label{eq:local_delay_def}
\end{equation}
In the longitudinal geometry, the photon-like branch stays essentially luminal, as the Euler--Heisenberg transverse refractive terms vanish when $\mathbf{k}\parallel\mathbf{B}$. As it was shown in Sec.\ref{transversal}, the medium in transverse geometry is birefringent, and the numerical continuation of Appendix~\ref{numerics} yields two photon-like branches, with group velocities calculated in eq.\eqref{vperpperp} and eq.\eqref{vperppara}. Therefore, evaluating eq.\eqref{eq:local_delay_def} for each branch gives
\begin{equation}
\Delta t_{L}^{\rm int}\simeq -2.5\times10^{-30}~\mathrm{s},
\end{equation}
for the longitudinal case, and
\begin{equation}
\Delta t_{T\perp}^{\rm int}\simeq 5.3\times10^{-12}~\mathrm{s},
\qquad
\Delta t_{T\parallel}^{\rm int}\simeq 9.3\times10^{-12}~\mathrm{s},
\label{eq:local_multimessenger_delays}
\end{equation}
for the two transverse branches.

These local values expose a strong anisotropy induced by the magnetized environment. The longitudinal delay is set merely by the residual neutrino lag over $30~\mathrm{km}$, while the transverse branches, refractively slowed, accumulate delays many orders of magnitude larger and split apart by the birefringence of the medium. The structure of these results is itself instructive: the transverse delays are set predominantly by the Euler--Heisenberg refractive response of the magnetized vacuum. The axion cloud reshapes the coupled spectrum through its density-dependent, time-dependent, and parity-odd contributions, generating the longitudinal circular splitting and adding axion-specific parity-odd structure, but it does not drive the leading transverse delay at the benchmark point. The two-branch transverse splitting in eq.~\eqref{eq:local_multimessenger_delays} is therefore a direct kinematic signature of the Euler--Heisenberg birefringence of the magnetized vacuum, with axion-induced corrections remaining subleading for the benchmark parameters.

Despite this theoretical anisotropy, all of these local delays remain microscopic. Even when combined with the full cosmological propagation baseline, no configuration approaches the $\sim 38~\mathrm{s}$ separation reported for the candidate event \cite{Abbasi_2024}. Local dispersion within the intervening pulsar-hosted environment cannot explain macroscopic GRB--neutrino offsets. Nor can a cumulative effect rescue this picture, since reproducing such an offset would require the line of sight to cross many magnetized environments, and the probability of even successive pulsar-scale encounters over cosmological distances is strongly suppressed (Appendix~\ref{sec:magnetar_model}).

The lasting result is therefore not an explanation of the observed offsets but a change of scale. The transverse delay obtained here exceeds, by many orders of magnitude, those found for diffuse astrophysical backgrounds such as homogeneous axion dark matter or the interstellar medium \cite{Masudepjc, Optical}. Strongly magnetized pulsar environments thus enhance the local dispersive response of the vacuum far beyond any diffuse medium, while the coexisting axion cloud imprints density-dependent, time-dependent, and parity-odd birefringent structure on top of this QED baseline.

\section{Polarization Observables: QED Baseline and Axion-Induced Circular Birefringence}
\label{sec:polarization}

The propagation operator derived in Sec.~\ref{theory} contains two
qualitatively distinct polarization effects. The parity-even
Euler--Heisenberg response determines the linear birefringence of the
magnetized vacuum, while the time-dependent axion background generates an
antisymmetric contribution that splits the two circular polarizations. We
first establish the hierarchy in the linear channel and then determine the
observable consequences of the parity-odd axion contribution.

\subsection{QED-dominated linear birefringence}
\label{sec:linear_qed}

For propagation perpendicular to the external magnetic field, the photon
polarization $A_{\parallel}$, whose electric field is parallel to
$\mathbf{B}_T$, mixes with the axion fluctuation. To display the hierarchy of
the diagonal terms in this channel, we supplement the local vacuum response
derived in Sec.~\ref{disprel} with the standard scalar plasma contribution.
In the short-wavelength approximation, following the canonical weak-mixing formalism \cite{Mixing1988}, the corresponding propagation equation is
\begin{equation}
i\frac{d}{ds}
\begin{pmatrix}
A_{\parallel}\\
a
\end{pmatrix}
=
\begin{pmatrix}
\Delta_{\parallel} & \Delta_{a\gamma}\\
\Delta_{a\gamma} & \Delta_a
\end{pmatrix}
\begin{pmatrix}
A_{\parallel}\\
a
\end{pmatrix},
\label{eq:mixing_matrix_linear}
\end{equation}
where
\begin{equation}
\Delta_{\parallel}
=
-\frac{\omega_{\rm pl}^{2}}{2\omega}
+
\Delta_{\parallel}^{\rm QED},
\qquad
\Delta_a
=
-\frac{m_a^{2}}{2\omega},
\qquad
\Delta_{a\gamma}
=
\frac{g_{a\gamma\gamma}B_T}{2}.
\label{eq:mixing_entries_linear}
\end{equation}
The QED diagonal term is related to the refractive-index shift of the
mixing photon mode by
\begin{equation}
\Delta_{\parallel}^{\rm QED}
=
\omega\,\delta n_{\parallel}^{\rm QED},
\qquad
\delta n_{\parallel}^{\rm QED}
=
28\epsilon B_T^{2},
\label{eq:qed_diagonal_linear}
\end{equation}
in agreement with the classical Euler--Heisenberg birefringent branch \cite{EH1936, Adler1971} obtained in
eq.~\eqref{eq:QED_index_parallel}.

A level crossing occurs when the background medium alters the photon's effective mass to perfectly match the axion mass, enabling highly efficient, resonant conversion between the two states. Away from this resonance, however, the perturbative-mixing condition applies:
\begin{equation}
\left|
\Delta_{a\gamma}
\right|
\ll
\left|
\Delta_{\parallel}-\Delta_a
\right|
\end{equation}
allowing the photon-like eigenvalue to be written as
\begin{equation}
\delta\Delta_{\gamma}^{\rm lin}
\simeq
\frac{\Delta_{a\gamma}^{2}}
{\Delta_{\parallel}-\Delta_a}.
\end{equation}
The associated refractive-index correction is therefore
\begin{equation}
\delta n_a^{\rm lin}
=
\frac{\delta\Delta_{\gamma}^{\rm lin}}{\omega}
\simeq
\frac{
\left(
g_{a\gamma\gamma}B_T
\right)^2
}{
2\left[
m_a^2-\omega_{\rm pl}^{2}
+
2\omega^{2}\delta n_{\parallel}^{\rm QED}
\right]
}.
\label{eq:dn_lin_full}
\end{equation}
The photon--axion splitting is thus fixed by the full diagonal response of
the medium rather than by the axion mass alone. Near a level crossing, the
matrix in eq.~\eqref{eq:mixing_matrix_linear} must instead be diagonalized
without expanding in $\Delta_{a\gamma}$.

At the benchmark point of Table~\ref{tab:benchmark}, the QED shift of the
mixing photon mode is
\begin{equation}
\delta n_{\parallel}^{\rm QED}
\simeq
9.3\times10^{-8}.
\end{equation}
To isolate the hierarchy between the vacuum contributions, we set
$\omega_{\rm pl}=0$ in the following numerical comparison. The QED term in
the diagonal splitting then satisfies
\begin{equation}
2\omega^2\delta n_{\parallel}^{\rm QED}
\simeq
1.9\times10^{-15}\,\mathrm{GeV}^{2},
\end{equation}
whereas
\begin{equation}
m_a^2
=
10^{-26}\,\mathrm{GeV}^{2}.
\end{equation}
Equation~\eqref{eq:dn_lin_full} consequently gives
\begin{equation}
\delta n_a^{\rm lin}
\simeq
1.0\times10^{-23},
\label{eq:linear_axion_benchmark}
\end{equation}
approximately sixteen orders of magnitude below the QED refractive shift.

The linear polarization eigenstates and their phase evolution are therefore
set by the Euler--Heisenberg response at the benchmark point. The
mixing-induced axion correction is too small to support a competitive
constraint on $g_{a\gamma\gamma}$ in this channel without a dedicated
polarization-transfer analysis of the magnetic and plasma profiles.

\subsection{Axion-induced circular birefringence}
\label{sec:circular}

The time dependence of the coherent axion field produces the antisymmetric
term proportional to
$\dot{\phi}_B\epsilon^{ijm}k_m$ in eq.~\eqref{Mtensor}. This structure is
absent from the parity-even Euler--Heisenberg response and splits the two
circular polarization eigenstates.

The effect is most transparent in the longitudinal geometry,
$\mathbf{k}\parallel\mathbf{B}$. Identifying the photon-like branches
$\omega_{1,2}$ of Sec.~\ref{disprel} with the two helicity eigenstates, their
dispersion relations can be written as
\begin{equation}
\omega_{\pm}^{2}
=
k^{2}
\mp
\left(\frac{
g_{a\gamma\gamma}\dot{\phi}_B
}{
c_1
} \right)k.
\label{eq:circular_branches}
\end{equation}
For fixed frequency and $\left|\frac{
g_{a\gamma\gamma}\dot{\phi}_B
}{
c_1
} \right|\ll\omega$, the corresponding wave numbers
are
\begin{equation}
k_{\pm}
\simeq
\omega
\pm
\frac{
g_{a\gamma\gamma}\dot{\phi}_B
}{
2c_1
},
\end{equation}
up to the convention used to label the two helicities. Their local
refractive-index splitting is therefore
\begin{equation}
\Delta n_{\rm circ}
\equiv
n_+-n_-
\simeq
\frac{
g_{a\gamma\gamma}\dot{\phi}_B
}{
c_1\omega
}.
\label{eq:circular_index_splitting}
\end{equation}
This contribution is parity-odd, linear in $g_{a\gamma\gamma}$, and directly
sensitive to the coherent axion background, reflecting the characteristic Carroll-Field-Jackiw chiral dispersion \cite{Carroll1990}.

A linearly polarized wave is a superposition of the two circular
eigenstates. Their relative phase rotates the polarization plane by
\begin{equation}
\Delta\alpha_{\rm odd}
=
\frac12
\int_{\rm path}
\left(
k_+-k_-
\right)ds.
\label{eq:rotation_integral}
\end{equation}
For a spacetime-dependent background, the field variation along the photon
trajectory is
\begin{equation}
\frac{d\phi_B}{ds}
=
\frac{\partial\phi_B}{\partial t}
+
\hat{\mathbf{k}}\cdot\nabla\phi_B.
\label{eq:axion_ray_derivative}
\end{equation}
The integrated rotation can then be expressed as
\begin{equation}
\Delta\alpha_{\rm odd}
=
\frac{g_{a\gamma\gamma}}{2}
\int_{\rm path}
\frac{1}{c_1}
\frac{d\phi_B}{ds}\,ds.
\end{equation}
At leading order in the weak-field expansion, $c_1\simeq1$, and hence
\begin{equation}
\Delta\alpha_{\rm odd}
\simeq
\frac{g_{a\gamma\gamma}}{2}
\left[
\phi_B(x_o)-\phi_B(x_e)
\right],
\label{eq:circ_rotation}
\end{equation}
where $x_e$ and $x_o$ denote the physical emission and observation events. The integrated rotation therefore depends only on the axion-field contrast between the endpoints of the ray, a foundational result for pseudoscalar-induced birefringence \cite{Harari1992}. For recent applications of this phase accumulation in oscillating dark matter backgrounds, see e.g., ref.~\cite{Fedderke2019}.

The local circular splitting may therefore remain nonzero throughout the
cloud, while the leading integrated rotation depends only on the axion-field
contrast between the endpoints of the ray. In the background-GRB transit
geometry of Sec.~\ref{timedelay}, both the source and the observer lie outside
the localized cloud,
\begin{equation}
\phi_B(x_e)
\simeq
\phi_B(x_o)
\simeq
0.
\end{equation}
The endpoint contribution then cancels,
\begin{equation}
\Delta\alpha_{\rm odd}^{\rm transit}
\simeq
0.
\label{eq:transit_rotation_cancellation}
\end{equation}
A complete vacuum-to-vacuum traversal thus samples a locally chiral medium
without acquiring a net polarization rotation at leading geometric-optics order \cite{Harari1992}.

A finite rotation requires a nonzero endpoint contrast. This occurs when the polarized source or the observer is embedded in the axion background, or when
the ray terminates within the cloud. Nonadiabatic boundary effects and
spatial mode conversion may also modify the endpoint result, but they lie
beyond the local homogeneous treatment adopted here.

\subsection{Endpoint contrast and local polarized emission}
\label{sec:circular_sensitivity}

For configurations with a nonzero endpoint contrast, eq.\eqref{eq:circ_rotation} defines a direct polarimetric target. The most natural astrophysical realization of this geometry occurs when the GRB is produced intrinsically by the pulsar, rather than acting as a background source transiting the line of sight.

This intrinsic emission scenario provides two major analytical advantages. First, the probability of the emitted photon encountering the dense axion cloud is identically unity, bypassing the severe statistical suppression for line-of-sight transits calculated in Appendix~\ref{sec:magnetar_model}. Second, because the emission event $x_e$ is embedded deep within the axion environment while the observer resides in the standard vacuum ($\phi_B(x_o) \simeq 0$), the geometric cancellation of eq.\eqref{eq:transit_rotation_cancellation} is strictly avoided. 

Crucially, the leading endpoint rotation depends on the value of the axion field at the physical endpoints of the photon trajectory, rather than on a photon--neutrino emission-time comparison. Unlike the multimessenger time-delay observable analyzed in Sec.~\ref{timedelay}, a polarimetric measurement does not require establishing a clean common emission time $t=0$ between different particles. It is therefore not affected by the same timing degeneracies that complicate the multimessenger observable, although its interpretation still depends on the polarized emission model, the magnetospheric plasma, the field geometry, and possible propagation effects inside the pulsar environment.

Since the axion field was defined in eq.\eqref{axion_field} as a coherent classical scalar field, the largest possible contrast between two endpoints sampling the same local
field amplitude is
\begin{equation}
\left|
\phi_B(x_o)-\phi_B(x_e)
\right|
\leq
2\phi_0.
\end{equation}
The corresponding rotation envelope is
\begin{equation}
\left|
\Delta\alpha_{\rm odd}
\right|_{\rm env}
\leq
g_{a\gamma\gamma}\phi_0
=
\frac{
g_{a\gamma\gamma}\sqrt{2\rho_a}
}{
m_a
}.
\label{eq:rotation_envelope}
\end{equation}
If only one endpoint lies within the cloud, the maximum contrast is
$\phi_0$, and the envelope is reduced by a factor of two.

Holding the local cloud density fixed at the bound-state benchmark value established for magnetospheric environments \cite{Noordhuis2024}, the maximal
two-endpoint envelope becomes
\begin{equation}
\left|
\Delta\alpha_{\rm odd}
\right|_{\rm env}
\lesssim
3.9\times10^{-3}
\left(
\frac{
g_{a\gamma\gamma}
}{
10^{-11}\,\mathrm{GeV}^{-1}
}
\right)
\left(
\frac{
\rho_a
}{
10^{22}\,\mathrm{GeV\,cm^{-3}}
}
\right)^{1/2}
\left(
\frac{
10^{-9}\,\mathrm{eV}
}{
m_a
}
\right)
\mathrm{rad}.
\label{eq:rotation_envelope_numeric}
\end{equation}
At $m_a=10^{-4}\,\mathrm{eV}$, the same expression gives
$|\Delta\alpha_{\rm odd}|_{\rm env}\lesssim3.9\times10^{-8}\,\mathrm{rad}$
for the benchmark coupling and density.

In the optimal endpoint configuration, a polarimeter sensitive to a rotation
$\Delta\alpha_{\rm min}$ would probe
\begin{equation}
g_{a\gamma\gamma}^{\rm sens}
\sim
\frac{
\Delta\alpha_{\rm min}m_a
}{
\sqrt{2\rho_a}
}.
\label{eq:conditional_sensitivity}
\end{equation}
For a single endpoint inside the cloud, the corresponding coupling is larger
by a factor of two. This is a conditional sensitivity estimate rather than a
bound for the vacuum-to-vacuum transit geometry. 

\subsection{Conditional sensitivity and the experimental landscape}
\label{sec:experimental_landscape}

To place the conditional endpoint sensitivity of eq.~\eqref{eq:conditional_sensitivity} in context, we compare it with current astrophysical bounds and projected laboratory sensitivities. Prompt GRB emission has been observed to exhibit substantial linear polarization in dedicated time-resolved analyses \cite{Nat.Astro,Chattopadhyay_2019}, motivating gamma-ray polarimetry as a relevant observational field. However, the endpoint benchmark derived here should not be interpreted as a constraint from those GRB data. We therefore adopt an illustrative polarimetric rotation sensitivity
\(\Delta\alpha_{\rm min}\simeq0.1\,\mathrm{rad}\), representative of the level of angular precision targeted by next-generation gamma-ray polarimetric analyses \cite{AMEGO,eASTROGAM2017}. For a favorable endpoint configuration in a canonical pulsar environment, with
\(\rho_a=10^{22}\,\mathrm{GeV\,cm^{-3}}\) and \(m_a=10^{-9}\,\mathrm{eV}\), this gives

\begin{equation}
    \boxed{g_{a\gamma\gamma}^{\rm sens} \lesssim 2.5\times10^{-10}\ \mathrm{GeV}^{-1}.}
    \label{eq:our_numerical_bound}
\end{equation}
This conditional result approaches the canonical coupling scale associated with the leading helioscope and stellar-cooling constraints, for which representative limits sit at order $g_{a\gamma\gamma} \sim 10^{-11}\text{--}10^{-10}\,\mathrm{GeV}^{-1}$ \cite{Irastorza2018,AxionLimits}. 

Crucially, the strength of this polarimetric sensitivity is highly mass-dependent. Because the instantaneous amplitude of the background axion field scales as $\phi_0 \propto 1/m_a$ for a fixed local dark matter density, lighter axions induce a significantly larger macroscopic rotation of the polarization plane. The mechanism is therefore optimally sensitive in the ultralight regime. Conversely, if the benchmark mass is shifted to $m_a = 10^{-4}\,\mathrm{eV}$, the local field amplitude is suppressed, shrinking the rotation envelope by five orders of magnitude. The corresponding sensitivity then degrades to $g_{a\gamma\gamma}^{\rm sens} \lesssim 2.5\times10^{-5}\,\mathrm{GeV}^{-1}$, entering a parameter space that is already robustly excluded by standard horizontal-branch star observations \cite{Ayala2014}.

This mass-dependent sensitivity must be interpreted carefully when comparing our framework with existing transient astrophysical bounds. The non-observation of a coincident gamma-ray flare from SN1987A \cite{Payez2015}, together with recent searches for ALP-induced gamma-ray emission from nearby pre-supernova stars \cite{Mittal2025}, constrains the coupling at the level of $g_{a\gamma\gamma} \lesssim 5.3\times10^{-12}\,\mathrm{GeV}^{-1}$ and $(0.13\text{--}1.26)\times10^{-11}\,\mathrm{GeV}^{-1}$, respectively; these are complemented by updated limits derived from the modeling of axion emission and magnetic conversion in such transient environments \cite{Fiorillo:2026sn}. In addition, recent polarimetric studies of the Crab pulsar have placed limits of $g_{a\gamma\gamma} \lesssim 1.5\times10^{-10}\,\mathrm{GeV}^{-1}$ for ultralight axions \cite{pulsar_polarimetry}. 

While these existing bounds are highly significant, they probe a qualitatively different physical regime and rely on distinct observational channels. In particular, they predominantly constrain axion-like particles through long-baseline intergalactic magnetic field conversions, pre-supernova emission, or continuous pulsar tracking. By contrast, the present result is rooted in localized, coherent circular birefringence in a pulsar-hosted axion cloud, and applies to configurations in which the physical endpoints of the photon trajectory sample different values of the axion background.

Viewed in this way, the significance of eq.~\eqref{eq:our_numerical_bound} is not that it supersedes existing limits in a universal sense, but that it identifies a targeted probe of extreme local environments. Under favorable endpoint conditions in canonical pulsar environments, the circular-birefringence channel defines a conditional sensitivity benchmark in the ultralight axion regime. This benchmark is complementary to laboratory and helioscope searches, rather than a replacement for them, and can be compared with the projected sensitivities of programs such as IAXO, MADMAX, and plasma haloscopes \cite{IAXO2019,Carenza2025,MADMAX,Lawson:2019plasma}. More broadly, it shows that highly magnetized pulsar environments can provide independent laboratories for testing axion electrodynamics through parity-odd birefringent effects.

\section{Conclusions}
\label{sec:conclusions}

Temporal offsets between GRB photons and high-energy neutrinos provide a useful probe of propagation effects in extreme astrophysical environments. In this work, we investigated whether such offsets could be generated by photon propagation through a dense axion cloud gravitationally bound to a magnetized compact object. To address this question, we constructed an effective local description based on the Euler--Heisenberg action extended by the axion sector and derived the corresponding dispersion relations in the presence of a strong magnetic background and an oscillating axion field.

The propagation problem contains two distinct ingredients. The Euler--Heisenberg sector provides the dominant parity-even birefringent response of the magnetized vacuum, while the coherent axion background adds mixing, explicit time dependence, and a parity-odd contribution to the local propagation operator. The resulting photon velocities depend on the orientation of the ray relative to the external magnetic field. For the most dispersive benchmark branch, obtained in the transverse configuration, we find
\begin{equation}
\Delta t_{T\parallel}^{\rm int}
\simeq
9.3\times10^{-12}~\mathrm{s},
\label{eq:conclusions_delay}
\end{equation}
for a traversal of the inner magnetized region. This delay is much larger than the values usually associated with diffuse astrophysical media, but it is still far below the $\mathcal{O}(1)$--$\mathcal{O}(10)\,\mathrm{s}$ offsets relevant for current multimessenger candidates. The local dispersive mechanism studied here therefore cannot account for macroscopic GRB--neutrino time separations.

The full benchmark time-of-flight estimate reinforces this conclusion. Once the local photon contribution is combined with the neutrino kinematic correction over the cosmological baseline, the total arrival-time difference remains microscopic and is not controlled by the axion-cloud transit. Higher-order QED corrections are not expected to change this conclusion qualitatively within the sub-Schwinger regime considered here, $B<B_{\mathrm{critical}}$. Likewise, replacing a diffuse background by a pulsar-hosted axion cloud enhances the local response, but not enough to challenge Lorentz-invariance-violation interpretations of macroscopic GRB offsets on time-delay grounds alone.

The same environment also contains an axion-specific polarimetric channel. The linear birefringence is dominated by the Euler--Heisenberg contribution for the benchmark parameters, so the linear polarization sector cannot be converted into a competitive bound on $g_{a\gamma\gamma}$ without a full polarization-transfer model. The genuinely axionic signature is instead the parity-odd circular birefringence generated by the oscillating background field. Its leading integrated effect is controlled by the difference of the axion field between the physical endpoints of the photon trajectory. Thus, a complete vacuum-to-vacuum traversal of a localized cloud gives no leading net endpoint rotation, even though the local circular splitting is nonzero inside the cloud.

For configurations in which polarized emission samples a nonzero axion-field contrast between endpoints, this circular channel defines a conditional local sensitivity. In the optimistic endpoint configuration considered in Sec.~\ref{sec:polarization}, the reference reach is
\begin{equation}
g_{a\gamma\gamma}
\lesssim
2.5\times10^{-10}\,\mathrm{GeV}^{-1},
\label{eq:conclusions_bound}
\end{equation}
for canonical pulsar fields, $\rho_a\simeq10^{22}\,\mathrm{GeV\,cm^{-3}}$, and ultralight axions near $m_a\sim10^{-9}\,\mathrm{eV}$. This number should not be read as a model-independent exclusion limit, nor as a constraint associated with the background-GRB transit geometry used in the time-delay calculation. It is a benchmark sensitivity for geometries in which the emission or detection endpoint lies inside the axion background, or more generally in which the endpoint values of $\phi_B$ differ.

The mass dependence of this polarimetric benchmark follows directly from the coherent-field amplitude, $\phi_0\propto \sqrt{\rho_a}/m_a$. At fixed local density, lighter axions therefore produce a larger endpoint phase, while the sensitivity rapidly degrades toward the upper end of the cloud-production mass window. This behavior distinguishes the circular-birefringence channel from the QED-dominated linear response and identifies the ultralight regime as the most favorable domain for this observable.

Taken together, these results show that pulsar-hosted axion clouds provide a controlled setting in which dispersive and birefringent effects can be compared within the same local effective theory. The time-delay signal remains too small to explain macroscopic multimessenger offsets, and the probability of repeated compact-object transits is negligible. The polarimetric channel is more promising conceptually, but only under endpoint conditions that avoid the leading cancellation of a complete localized-cloud traversal. A realistic constraint would therefore require a dedicated model of the polarized source, the emission region, the magnetospheric plasma, and the axion-cloud profile. Within these limitations, compact-object environments remain useful laboratories for axion electrodynamics, especially through parity-odd birefringent signatures that are absent from the Euler--Heisenberg vacuum.

\appendix

\section{Numerical Determination of the Transverse Branches}
\label{numerics}

The longitudinal photon branches discussed in Sec.~\ref{disprel} are obtained
analytically. In the transverse configuration, we determine numerically the
two photon-like branches and the axion-like branch of the coupled propagation
system. The calculation is local: within the WKB approximation, the
background quantities are held fixed while solving the dispersion relation.

\paragraph{Regular coupled system.}

The effective photon determinant in eq.~\eqref{DetMperp} contains the axion
denominator
\begin{equation}
\mathcal{D}_a
=
k^2+m_a^2-\omega^2,
\end{equation}
because the axion fluctuation has been eliminated algebraically. For the
numerical calculation, we retain this fluctuation and solve the regular
system in the basis $(e_y,e_z,\chi_a)$, where $\chi_a$ is a rescaled axion
amplitude. For
\begin{equation}
\mathbf{k}=k\,\hat{\mathbf{x}},
\qquad
\mathbf{B}=B_z\,\hat{\mathbf{z}},
\qquad
\mathbf{E}=E_z\,\hat{\mathbf{z}},
\end{equation}
the coupled operator is
\begin{equation}
\mathbb{M}_\perp
=
\begin{pmatrix}
A_0 & Z_0 & U_y\\
Z_0^{*} & C_0 & U_z\\
U_y & U_z & -\mathcal{D}_a
\end{pmatrix},
\label{eq:regular_coupled_matrix}
\end{equation}
with
\begin{align}
A_0
&=
\omega^2-k^2
+
\frac{d_1}{c_1}k^2B_z^2
+
\frac{d_2}{c_1}k^2E_z^2,
\\
C_0
&=
\omega^2-k^2
+
\frac{d_2}{c_1}\omega^2B_z^2
+
\frac{d_1}{c_1}\omega^2E_z^2,
\\
Z_0
&=
\frac{d_2-d_1}{c_1}\,
\omega k B_zE_z
-
i\frac{g_{a \gamma\gamma}\dot{\phi}_B}{c_1}k,
\\
U_y
&=
\frac{g_{a \gamma\gamma}kE_z}{\sqrt{c_1}},
\qquad
U_z
=
\frac{g_{a \gamma\gamma}\omega B_z}{\sqrt{c_1}}.
\end{align}
For real $\omega$, this operator is Hermitian.

The numerical roots are obtained from the dimensionless secular function
\begin{equation}
\mathcal{F}(\omega,k)
\equiv
\frac{\det\mathbb{M}_\perp(\omega,k)}{k^6}
=
0.
\label{eq:regular_secular_equation}
\end{equation}
The normalization by $k^6$ improves the numerical conditioning without
changing the roots.

Eliminating $\chi_a$ through the Schur complement reproduces the effective
photon determinant of eq.~\eqref{DetMperp}. More precisely,
\begin{equation}
\det\mathbb{M}_\perp
=
-\mathcal{D}_a\,
\frac{\det M_\perp}{\omega^2}.
\label{eq:schur_determinant_relation}
\end{equation}
The regular formulation therefore replaces the pole of the reduced photon
operator by an explicit axion-like branch.

\paragraph{Reference states and branch continuation.}

The three branches are initialized at vanishing axion coupling,
\begin{equation}
g_{a \gamma\gamma}=0,
\qquad
E_z=0,
\end{equation}
while the external magnetic field is retained. Their positive-frequency
roots are
\begin{equation}
\omega_{\perp}^{(0)}
=
k
\sqrt{
1-\frac{d_1B_z^2}{c_1^{(0)}}
},
\qquad
\omega_{\parallel}^{(0)}
=
\frac{k}{
\sqrt{
1+\dfrac{d_2B_z^2}{c_1^{(0)}}
}
},
\qquad
\omega_a^{(0)}
=
\sqrt{k^2+m_a^2},
\label{eq:numerical_reference_roots}
\end{equation}
where
\begin{equation}
c_1^{(0)}
=
1+16\epsilon
\left(
\vec{E}^{\,2}-\vec{B}^{\,2}
\right).
\end{equation}
The corresponding eigenvectors are initially aligned with
$e_y$, $e_z$, and $\chi_a$, respectively.

The interacting system is reached through
\begin{equation}
g_\lambda=\lambda g,
\qquad
E_{z,\lambda}=g_\lambda B_z\phi_B,
\qquad
0\leq\lambda\leq1.
\label{eq:numerical_homotopy}
\end{equation}
All coefficients, including $c_1$, are reevaluated at each value of
$\lambda$.

A first-order estimate for the root at the next continuation step follows
from
\begin{equation}
\frac{d\omega}{d\lambda}
=
-
\frac{\partial_\lambda\mathcal{F}}
{\partial_\omega\mathcal{F}}.
\label{eq:lambda_predictor}
\end{equation}
This estimate and the preceding root are used as initial values for
high-precision Newton and secant iterations. The continuation begins with
40 steps between $\lambda=0$ and $\lambda=1$. If a step fails, its size is
halved and subsequently restored after successful iterations.

At each step, a normalized null vector $\mathbf{u}_j$ is constructed from
\begin{equation}
\mathbb{M}_\perp(\omega_j,k)\mathbf{u}_j=0.
\end{equation}
The roots are assigned to the three branches by maximizing the overlap
\begin{equation}
\mathcal{O}_{ij}
=
\left|
\mathbf{u}_i^\dagger(\lambda)
\mathbf{u}_j(\lambda+\Delta\lambda)
\right|.
\label{eq:eigenvector_overlap}
\end{equation}

\paragraph{Precision and group velocity.}

The calculation is performed with 100 significant digits. Accepted roots
have positive real frequency, relative imaginary part below $10^{-45}$, and
dimensionless residual
\begin{equation}
|\mathcal{F}(\omega,k)|<10^{-50}.
\end{equation}
No restriction on the group velocity or on the distance from the vacuum
light cone is imposed when assigning the branches.

For a simple root $\omega_j(k)$, the group velocity is evaluated through
implicit differentiation,
\begin{equation}
v_{g,j}
=
\frac{d\omega_j}{dk}
=
-
\left.
\frac{\partial_k\mathcal{F}}
{\partial_\omega\mathcal{F}}
\right|_{\omega=\omega_j(k)}.
\label{eq:vgimplicit}
\end{equation}
The derivatives are computed with arbitrary-precision numerical
differentiation. As a stability check, they are also evaluated with centered
finite differences using
\begin{equation}
h_\omega
=
\eta\max(|\omega|,m_a),
\qquad
h_k
=
\eta |k|,
\end{equation}
for
\begin{equation}
\eta
=
10^{-12},
10^{-16},
10^{-20},
10^{-24},
10^{-28}.
\end{equation}

It should be emphasized that in a general anisotropic medium the group velocity must be evaluated via the gradient, $\vec{v}_g=\nabla_{\vec{k}}\,\omega$, since the group and phase velocities are not in general collinear. The analysis performed here, however, targets precisely the two canonical configurations, longitudinal ($\vec{k}\parallel\vec{B}$) and transverse ($\vec{k}\perp\vec{B}$), in which the background fields define symmetry axes of the problem. Because the angular dependence of $\det M$ enters only through structures such as $(\vec{k}\cdot\vec{B})^{2}$,
$(\vec{k}\times\vec{B})^{2}$, and $\dot{\phi}_B^{2}\,\vec{k}^{2}$, the momentum components orthogonal to the respective propagation axis appear strictly through quadratic or higher even powers, as can be verified explicitly in eqs.~\eqref{detM0} and~\eqref{DetMperp}. Consequently, the components of $\nabla_{\vec{k}}\det M$ orthogonal to the propagation direction vanish identically when evaluated on these symmetry axes, by parity, and the three-dimensional gradient collapses to its single component along the propagation direction. This guarantees that the one-dimensional derivative of eq.~\eqref{eq:vgimplicit} yields the exact magnitude of the full group-velocity vector for the canonical modes considered here.

\paragraph{Validation.}

The implementation is subjected to the following checks before the
benchmark results are evaluated:

\begin{enumerate}

\item The determinant obtained directly from the photon propagation tensor
in eq.~\eqref{Mtensor} agrees with the compact expression in
eq.~\eqref{DetMperp} after removal of the trivial factor $\omega^2$.

\item The determinant of the regular coupled matrix satisfies the
Schur-complement identity in eq.~\eqref{eq:schur_determinant_relation}.

\item In the vacuum limit, the two photon roots reduce to $\omega=k$, while
the third root reduces to $\omega=\sqrt{k^2+m_a^2}$.

\item For $g=E_z=0$ and $B_z\neq0$, the two photon branches reproduce the
Euler--Heisenberg results in
eqs.~\eqref{eq:QED_index_perpendicular} and
\eqref{eq:QED_index_parallel}.

\item Reversing the axion phase,
$\theta_a\rightarrow-\theta_a$, leaves the three eigenfrequencies unchanged,
as required by the dependence of the secular equation on
$\dot{\phi}_B^{\,2}$.

\item The final roots and group velocities remain stable under variations of
the continuation step, working precision, and finite-difference scale.

\end{enumerate}

The resulting continuation identifies the two photon-like branches and the
axion-like branch by their evolution from the decoupled spectrum, rather
than by their proximity to the vacuum light cone.

\section{Conservative Estimate of the Pulsar-Environment Encounter Probability}
\label{sec:magnetar_model}

In Sec.~\ref{timedelay}, we showed that a single traversal through a pulsar-hosted axion cloud produces at most a picosecond-scale local delay. As discussed, the delay is evaluated for a photon from a background GRB transiting the environment, which fixes a clean common emission time for the photon--neutrino comparison. A natural question is then whether a macroscopic offset could arise from the cumulative effect of many such encounters along the line of sight.

The purpose of this appendix is to show that this possibility is statistically negligible. To do so, we construct a simple order-of-magnitude estimate for the expected number of encounters within a deliberately conservative Galactic benchmark. We model the active pulsar population by an average number density and assign a generous effective geometrical cross section to the surrounding environment to intentionally maximize the intersection probability.

For a population with average number density $n_{\rm psr}$ and an effective geometrical cross section $\sigma_{\rm env}$, the expected number of encounters along a path of length $D$ is
\begin{equation}
N_{\rm int} = n_{\rm psr}\,\sigma_{\rm env}\,D.
\label{eq:mean_encounters}
\end{equation}
In the dilute limit where $N_{\rm int} \ll 1$, this expected value directly approximates the probability of a single line-of-sight encounter, $P \simeq N_{\rm int}$.

To convert the underlying Galactic population into an average number density, we restrict the estimate to active canonical pulsars, since these are the objects for which the cloud-production mechanism and the benchmark magnetic fields used in this work are assumed. Population synthesis models, calibrated against large-scale radio observations, consistently constrain this active population size \cite{Vranesevic_2004,Xue_2023}. We adopt
\begin{equation}
N_{\rm total} = 10^5
\end{equation}
as a generous, observationally grounded estimate of the relevant active pulsars.

We model the Milky Way as a cylindrical disk with radius $R_{\rm MW} = 15~\mathrm{kpc}$ and thickness $h_{\rm MW} = 0.3~\mathrm{kpc}$, consistent with characteristic Galactic dimensions \cite{BlandHawthorn2016}. The effective volume is $V_{\rm MW} = \pi R_{\rm MW}^{2} h_{\rm MW}$, yielding a mean pulsar number density of
\begin{equation}
n_{\rm psr} = \frac{N_{\rm total}}{V_{\rm MW}} \simeq 4.7 \times 10^{2}~\mathrm{kpc}^{-3}.
\label{eq:mean_pulsar_density}
\end{equation}

The effective cross section used here must be distinguished from the dispersive length adopted in Sec.~\ref{timedelay}. A precise evaluation of the encounter probability would require integrating over the exact radial dependence of the dispersive effect, which is driven primarily by the rapid dipole decay ($1/r^3$) of the magnetic field away from the star. 

These profiles are not needed to establish a robust geometrical upper bound. Instead of modeling the continuous radial decay, we define a deliberately oversized outer envelope,
\begin{equation}
R_{\rm env} = 5 \times 10^4~\mathrm{km},
\end{equation}
and assign a corresponding cross section
\begin{equation}
\sigma_{\rm env} = \pi R_{\rm env}^{2}.
\label{eq:environment_cross_section}
\end{equation}
This radius is chosen solely to overestimate the effective target area. It treats the entire extended magnetosphere as a viable interaction target, bypassing the need for a detailed radial decay model and enforcing a worst-case scenario for the encounter probability.

For a representative Galactic path, $D_{\rm Gal} = 15~\mathrm{kpc}$, eq.~\eqref{eq:mean_encounters} gives
\begin{equation}
N_{\rm int} = n_{\rm psr}\pi R_{\rm env}^{2}D_{\rm Gal} \simeq 5.8 \times 10^{-20}.
\label{eq:mean_encounters_result}
\end{equation}
Even with this generously enlarged geometrical envelope, the chance of a single line-of-sight encounter is exceedingly unlikely. For comparison, the expected number of encounters for directly crossing the inner region of radius $R_{\rm int} = 15~\mathrm{km}$, where the strong benchmark fields actually reside, would be $N_{\rm int}^{\rm inner} \simeq 5.2 \times 10^{-27}$. The larger value in eq.~\eqref{eq:mean_encounters_result} should therefore be regarded as a highly conservative upper limit.

Because the expected number of single encounters is already of order $10^{-20}$, the probability of two or more independent encounters along the same trajectory scales as $N_{\rm int}^{2} \sim \mathcal{O}(10^{-40})$. Repeated crossings are decisively suppressed, meaning a macroscopic GRB--neutrino offset cannot be assembled cumulatively from successive pulsar crossings under any circumstances.

The transit configuration used in Sec.~\ref{timedelay} should consequently be interpreted as a controlled propagation benchmark rather than as a likely configuration for an observed GRB. Within this simplified population model, active canonical pulsars do not form an effectively diffuse dispersive medium along Galactic lines of sight. Successive crossings of ordinary pulsar-hosted axion clouds cannot occur frequently enough to accumulate into an observable multimessenger time delay, such as those discussed in ref.~\cite{Abbasi_2024}.

\acknowledgments

We are much grateful to Iver Brevik, Markku Oksanen, Anca Tureanu and Edoardo Vitagliano for fruitful discussions and suggestions, as well as to the two anonymous referees whose insightful questions and specific suggestions, have largely improved the work and its presentation. B.~A.~Couto e Silva acknowledges support from the Funda\c{c}\~ao de Amparo \`a Pesquisa do Estado de Minas Gerais (FAPEMIG) and the Coordena\c{c}\~ao de Aperfei\c{c}oamento de Pessoal de N\'ivel Superior (CAPES). B.~L.~S\'anchez-Vega acknowledges support from the Conselho Nacional de Desenvolvimento Cient\'ifico e Tecnol\'ogico (CNPq) through Grant No.~311699/2020-0.

\bibliographystyle{utphys.bst} 
\bibliography{biblio.bib}

@article{Fireball1,
  title = {{High Energy Neutrinos from Cosmological Gamma-Ray Burst Fireballs}},
  author = {Waxman, Eli and Bahcall, John},
  journal = {Phys. Rev. Lett.},
  volume = {78},
  issue = {12},
  pages = {2292--2295},
  year = {1997},
  publisher = {American Physical Society},
  doi = {10.1103/PhysRevLett.78.2292},

}

@article{Fireball2,
  title = {{Neutrino Emission from Gamma-Ray Burst Fireballs, Revised}},
  author = {H\"ummer, Svenja and Baerwald, Philipp and Winter, Walter},
  journal = {Phys. Rev. Lett.},
  volume = {108},
  issue = {23},
  pages = {231101},
  year = {2012},
  publisher = {American Physical Society},
  doi = {10.1103/PhysRevLett.108.231101},
}

@article{Berezinsky1969,
  author  = {Berezinsky, V. S. and Zatsepin, G. T.},
  title   = {Cosmic rays at ultra high energies (neutrino?)},
  journal = {Physics Letters B},
  volume  = {28},
  number  = {6},
  pages   = {423--424},
  year    = {1969},
  doi     = {10.1016/0370-2693(69)90341-4}
}

@article{Stecker1978,
    author = "Stecker, F. W.",
    title = "{{Diffuse Fluxes of Cosmic High-Energy Neutrinos}}",
    reportNumber = "NASA-TM-79609",
    doi = "10.1086/156919",
    journal = "Astrophys. J.",
    volume = "228",
    pages = "919--927",
    year = "1979"
}

@article{Waxman1996,
  author  = {Waxman, Eli and Coppi, Paolo},
  title   = {{Delayed GeV--TeV Photons from Gamma-Ray Bursts Producing High-Energy Cosmic Rays}},
  journal = {The Astrophysical Journal Letters},
  year    = {1996},
  volume  = {464},
  pages   = {L75--L78},
  doi     = {10.1086/310090}
}

@article{MiraldaEscude1996,
  author = {Miralda-Escud{\'e}, Jordi and Waxman, Eli},
  title = {{Signatures of the Origin of High-Energy Cosmic Rays in Cosmological Gamma-Ray Bursts}},
  journal = {Astrophys. J.},
  volume = {462},
  pages = {L59--L62},
  year = {1996},
  doi = {10.1086/310042},
  eprint = {astro-ph/9601012},
  archivePrefix = {arXiv}
}

@article{Murase2006,
  author  = {Murase, Kohta and Ioka, Kunihito and Nagataki, Shigehiro and Nakamura, Takashi},
  title   = {{High-Energy Neutrinos and Cosmic Rays from Low-Luminosity Gamma-Ray Bursts?}},
  journal = {The Astrophysical Journal Letters},
  year    = {2006},
  volume  = {651},
  number  = {1},
  pages   = {L5--L8},
  doi     = {10.1086/509323},
  eprint  = {astro-ph/0607104},
  archivePrefix = {arXiv}
}

@article{Bustamante2015,
  author = {Bustamante, Mauricio and Baerwald, Philipp and Murase, Kohta and Winter, Walter},
  title = {Neutrino and cosmic-ray emission from multiple internal shocks in gamma-ray bursts},
  journal = {Nature Commun.},
  volume = {6},
  pages = {6783},
  year = {2015},
  doi = {10.1038/ncomms7783},
  eprint = {1409.2874},
  archivePrefix = {arXiv},
  primaryClass = {astro-ph.HE}
}

@article{Adler1971,
  author = {Adler, Stephen L.},
  title = {Photon splitting and photon dispersion in a strong magnetic field},
  journal = {Annals Phys.},
  volume = {67},
  number = {2},
  pages = {599--647},
  year = {1971},
  doi = {10.1016/0003-4916(71)90154-0}
}

@article{Tsai1974,
  author = {Tsai, Wu-Yang and Erber, Thomas},
  title = {Photon Pair Creation in Intense Magnetic Fields},
  journal = {Phys. Rev. D},
  volume = {10},
  pages = {492-499},
  year = {1974},
  doi = {10.1103/PhysRevD.10.492}
}

@article{Karbstein2013,
  author = {Karbstein, Felix},
  title = {Photon polarization tensor in a homogeneous magnetic or electric field},
  journal = {Phys. Rev. D},
  volume = {88},
  pages = {085033},
  year = {2013},
  doi = {10.1103/PhysRevD.88.085033},
  eprint = {1308.6184},
  archivePrefix = {arXiv},
  primaryClass = {hep-th}
}

@article{BlandHawthorn2016,
  author = {Bland-Hawthorn, Joss and Gerhard, Ortwin},
  title = {{The Galaxy in Context: Structural, Kinematic, and Integrated Properties}},
  journal = {Annual Review of Astronomy and Astrophysics},
  volume = {54},
  pages = {529-596},
  year = {2016},
  doi = {10.1146/annurev-astro-081915-023441}
}

@article{Noordhuis2024,
  author = {Noordhuis, Dion and Prabhu, Anirudh and Weniger, Christoph and Witte, Samuel J.},
  title = {Axion Clouds around Neutron Stars},
  journal = {Physical Review X},
  volume = {14},
  pages = {041015},
  year = {2024},
  doi = {10.1103/PhysRevX.14.041015}
}

@article{KM3NeT,
      author         = "{The KM3NeT Collaboration}",
      title          = "{Observation of an ultra-high-energy cosmic neutrino with KM3NeT}",
      journal        = "Nature",
      volume         = "638",
      year           = "2025",
      pages          = "376--382",
      doi            = "10.1038/s41586-024-08543-1",
}

@article{Abbasi_2024,
doi = {10.3847/1538-4357/ad220b},
year = {2024},
publisher = {The American Astronomical Society},
volume = {964},
number = {2},
pages = {126},
author = {Abbasi, R. and Ackermann, M. and Adams, J. and Agarwalla, S. K. and Aguilar, J. A. and Ahlers, M. and Alameddine, J. M. and Amin, N. M. and Andeen, K. and Anton, G. and Argüelles, C. and Ashida, Y. and Athanasiadou, S. and Ausborm, L. and Axani, S. N. and Bai, X. and V., A. Balagopal and Baricevic, M. and Barwick, S. W. and Basu, V. and Bay, R. and Beatty, J. J. and Becker Tjus, J. and Beise, J. and Bellenghi, C. and Benning, C. and BenZvi, S. and Berley, D. and Bernardini, E. and Besson, D. Z. and Blaufuss, E. and Blot, S. and Bontempo, F. and Book, J. Y. and Boscolo Meneguolo, C. and Böser, S. and Botner, O. and Böttcher, J. and Braun, J. and Brinson, B. and Brostean-Kaiser, J. and Brusa, L. and Burley, R. T. and Busse, R. S. and Butterfield, D. and Campana, M. A. and Carloni, K. and Carnie-Bronca, E. G. and Chattopadhyay, S. and Chau, N. and Chen, C. and Chen, Z. and Chirkin, D. and Choi, S. and Clark, B. A. and Coleman, A. and Collin, G. H. and Connolly, A. and Conrad, J. M. and Coppin, P. and Correa, P. and Cowen, D. F. and Dave, P. and De Clercq, C. and DeLaunay, J. J. and Delgado, D. and Deng, S. and Deoskar, K. and Desai, A. and Desiati, P. and de Vries, K. D. and de Wasseige, G. and DeYoung, T. and Diaz, A. and Díaz-Vélez, J. C. and Dittmer, M. and Domi, A. and Dujmovic, H. and DuVernois, M. A. and Ehrhardt, T. and Eimer, A. and Eller, P. and Ellinger, E. and El Mentawi, S. and Elsässer, D. and Engel, R. and Erpenbeck, H. and Evans, J. and Evenson, P. A. and Fan, K. L. and Fang, K. and Farrag, K. and Fazely, A. R. and Fedynitch, A. and Feigl, N. and Fiedlschuster, S. and Finley, C. and Fischer, L. and Fox, D. and Franckowiak, A. and Fürst, P. and Gallagher, J. and Ganster, E. and Garcia, A. and Gerhardt, L. and Ghadimi, A. and Glaser, C. and Glauch, T. and Glüsenkamp, T. and Gonzalez, J. G. and Grant, D. and Gray, S. J. and Gries, O. and Griffin, S. and Griswold, S. and Groth, K. M. and Günther, C. and Gutjahr, P. and Ha, C. and Haack, C. and Hallgren, A. and Halliday, R. and Halve, L. and Halzen, F. and Hamdaoui, H. and Ha Minh, M. and Handt, M. and Hanson, K. and Hardin, J. and Harnisch, A. A. and Hatch, P. and Haungs, A. and Häußler, J. and Helbing, K. and Hellrung, J. and Hermannsgabner, J. and Heuermann, L. and Heyer, N. and Hickford, S. and Hidvegi, A. and Hill, C. and Hill, G. C. and Hoffman, K. D. and Hori, S. and Hoshina, K. and Hou, W. and Huber, T. and Hultqvist, K. and Hünnefeld, M. and Hussain, R. and Hymon, K. and In, S. and Ishihara, A. and Jacquart, M. and Janik, O. and Jansson, M. and Japaridze, G. S. and Jeong, M. and Jin, M. and Jones, B. J. P. and Kamp, N. and Kang, D. and Kang, W. and Kang, X. and Kappes, A. and Kappesser, D. and Kardum, L. and Karg, T. and Karl, M. and Karle, A. and Katil, A. and Katz, U. and Kauer, M. and Kelley, J. L. and Khatee Zathul, A. and Kheirandish, A. and Kiryluk, J. and Klein, S. R. and Kochocki, A. and Koirala, R. and Kolanoski, H. and Kontrimas, T. and Köpke, L. and Kopper, C. and Koskinen, D. J. and Koundal, P. and Kovacevich, M. and Kowalski, M. and Kozynets, T. and Krishnamoorthi, J. and Kruiswijk, K. and Krupczak, E. and Kumar, A. and Kun, E. and Kurahashi, N. and Lad, N. and Lagunas Gualda, C. and Lamoureux, M. and Larson, M. J. and Latseva, S. and Lauber, F. and Lazar, J. P. and Lee, J. W. and DeHolton, K. Leonard and Leszczyńska, A. and Lincetto, M. and Liu, Y. and Liubarska, M. and Lohfink, E. and Love, C. and Lozano Mariscal, C. J. and Lu, L. and Lucarelli, F. and Luszczak, W. and Lyu, Y. and Madsen, J. and Magnus, E. and Mahn, K. B. M. and Makino, Y. and Manao, E. and Mancina, S. and Marie Sainte, W. and Mariş, I. C. and Marka, S. and Marka, Z. and Marsee, M. and Martinez-Soler, I. and Maruyama, R. and Mayhew, F. and McElroy, T. and McNally, F. and Mead, J. V. and Meagher, K. and Mechbal, S. and Medina, A. and Meier, M. and Merckx, Y. and Merten, L. and Micallef, J. and Mitchell, J. and Montaruli, T. and Moore, R. W. and Morii, Y. and Morse, R. and Moulai, M. and Mukherjee, T. and Naab, R. and Nagai, R. and Nakos, M. and Naumann, U. and Necker, J. and Negi, A. and Neumann, M. and Niederhausen, H. and Nisa, M. U. and Noell, A. and Novikov, A. and Nowicki, S. C. and Obertacke Pollmann, A. and O’Dell, V. and Oeyen, B. and Olivas, A. and Orsoe, R. and Osborn, J. and O’Sullivan, E. and Pandya, H. and Park, N. and Parker, G. K. and Paudel, E. N. and Paul, L. and Pérez de los Heros, C. and Peterson, J. and Philippen, S. and Pizzuto, A. and Plum, M. and Pontén, A. and Popovych, Y. and Prado Rodriguez, M. and Pries, B. and Procter-Murphy, R. and Przybylski, G. T. and Raab, C. and Rack-Helleis, J. and Rawlins, K. and Rechav, Z. and Rehman, A. and Reichherzer, P. and Resconi, E. and Reusch, S. and Rhode, W. and Riedel, B. and Rifaie, A. and Roberts, E. J. and Robertson, S. and Rodan, S. and Roellinghoff, G. and Rongen, M. and Rosted, A. and Rott, C. and Ruhe, T. and Ruohan, L. and Ryckbosch, D. and Safa, I. and Saffer, J. and Salazar-Gallegos, D. and Sampathkumar, P. and Sanchez Herrera, S. E. and Sandrock, A. and Santander, M. and Sarkar, S. and Sarkar, S. and Savelberg, J. and Savina, P. and Schaufel, M. and Schieler, H. and Schindler, S. and Schlickmann, L. and Schlüter, B. and Schlüter, F. and Schmeisser, N. and Schmidt, T. and Schneider, J. and Schröder, F. G. and Schumacher, L. and Sclafani, S. and Seckel, D. and Seikh, M. and Seunarine, S. and Shah, R. and Shefali, S. and Shimizu, N. and Silva, C. and Silva, M. and Skrzypek, B. and Smithers, B. and Snihur, R. and Soedingrekso, J. and Søgaard, A. and Soldin, D. and Soldin, P. and Sommani, G. and Spannfellner, C. and Spiczak, G. M. and Spiering, C. and Stamatikos, M. and Stanev, T. and Stezelberger, T. and Stürwald, T. and Stuttard, T. and Sullivan, G. W. and Taboada, I. and Ter-Antonyan, S. and Thiesmeyer, M. and Thompson, W. G. and Thwaites, J. and Tilav, S. and Tollefson, K. and Tönnis, C. and Toscano, S. and Tosi, D. and Trettin, A. and Tung, C. F. and Turcotte, R. and Twagirayezu, J. P. and Unland Elorrieta, M. A. and Upadhyay, A. K. and Upshaw, K. and Vaidyanathan, A. and Valtonen-Mattila, N. and Vandenbroucke, J. and van Eijndhoven, N. and Vannerom, D. and van Santen, J. and Vara, J. and Veitch-Michaelis, J. and Venugopal, M. and Vereecken, M. and Verpoest, S. and Veske, D. and Vijai, A. and Walck, C. and Wang, Y. and Weaver, C. and Weigel, P. and Weindl, A. and Weldert, J. and Wen, A. Y. and Wendt, C. and Werthebach, J. and Weyrauch, M. and Whitehorn, N. and Wiebusch, C. H. and Williams, D. R. and Witthaus, L. and Wolf, A. and Wolf, M. and Wrede, G. and Xu, X. W. and Yanez, J. P. and Yildizci, E. and Yoshida, S. and Young, R. and Yu, S. and Yuan, T. and Zhang, Z. and Zhelnin, P. and Zilberman, P. and Zimmerman, M. and IceCube Collaboration},
title = {{Search for 10–1000 GeV Neutrinos from Gamma-Ray Bursts with IceCube}},
journal = {The Astrophysical Journal},

}

@article{Antaresmuon,
	author = {{Adrián-Martínez, S.} and {Albert, A.} and {Al Samarai, I.} and {André, M.} and {Anghinolfi, M.} and {Anton, G.} and {Anvar, S.} and {Ardid, M.} and {Astraatmadja, T.} and {Aubert, J.-J.} and {Baret, B.} and {Barrios-Marti, J.} and {Basa, S.} and {Bertin, V.} and {Biagi, S.} and {Bigongiari, C.} and {Bogazzi, C.} and {Bouhou, B.} and {Bouwhuis, M. C.} and {Brunner, J.} and {Busto, J.} and {Capone, A.} and {Caramete, L.} and {Cârloganu, C.} and {Carr, J.} and {Cecchini, S.} and {Charif, Z.} and {Charvis, Ph.} and {Chiarusi, T.} and {Circella, M.} and {Classen, F.} and {Coniglione, R.} and {Core, L.} and {Costantini, H.} and {Coyle, P.} and {Creusot, A.} and {Curtil, C.} and {De Bonis, G.} and {Dekeyser, I.} and {Deschamps, A.} and {Distefano, C.} and {Donzaud, C.} and {Dornic, D.} and {Dorosti, Q.} and {Drouhin, D.} and {Dumas, A.} and {Eberl, T.} and {Emanuele, U.} and {Enzenhöfer, A.} and {Ernenwein, J.-P.} and {Escoffier, S.} and {Fehn, K.} and {Fermani, P.} and {Flaminio, V.} and {Folger, F.} and {Fritsch, U.} and {Fusco, L. A.} and {Galatà, S.} and {Gay, P.} and {Geißelsöder, S.} and {Geyer, K.} and {Giacomelli, G.} and {Giordano, V.} and {Gleixner, A.} and {Gómez-González, J. P.} and {Graf, K.} and {Guillard, G.} and {van Haren, H.} and {Heijboer, A. J.} and {Hello, Y.} and {Hernández-Rey, J. J.} and {Herold, B.} and {Hößl, J.} and {James, C. W.} and {de Jong, M.} and {Kadler, M.} and {Kalekin, O.} and {Kappes, A.} and {Katz, U.} and {Kooijman, P.} and {Kouchner, A.} and {Kreykenbohm, I.} and {Kulikovskiy, V.} and {Lahmann, R.} and {Lambard, E.} and {Lambard, G.} and {Larosa, G.} and {Lefèvre, D.} and {Leonora, E.} and {Lo Presti, D.} and {Loehner, H.} and {Loucatos, S.} and {Louis, F.} and {Mangano, S.} and {Marcelin, M.} and {Margiotta, A.} and {Martínez-Mora, J. A.} and {Martini, S.} and {Michael, T.} and {Montaruli, T.} and {Morganti, M.} and {Müller, C.} and {Neff, M.} and {Nezri, E.} and {Palioselitis, D.} and {Păvălaş, G. E.} and {Perrina, C.} and {Piattelli, P.} and {Popa, V.} and {Pradier, T.} and {Racca, C.} and {Riccobene, G.} and {Richter, R.} and {Rivière, C.} and {Robert, A.} and {Roensch, K.} and {Rostovtsev, A.} and {Samtleben, D. F. E.} and {Sanguineti, M.} and {Schmid, J.} and {Schnabel, J.} and {Schulte, S.} and {Schüssler, F.} and {Seitz, T.} and {Shanidze, R.} and {Sieger, C.} and {Simeone, F.} and {Spies, A.} and {Spurio, M.} and {Steijger, J. J. M.} and {Stolarczyk, Th.} and {Sánchez-Losa, A.} and {Taiuti, M.} and {Tamburini, C.} and {Tayalati, Y.} and {Trovato, A.} and {Vallage, B.} and {Vallée, C.} and {Van Elewyck, V.} and {Vernin, P.} and {Visser, E.} and {Wagner, S.} and {Wilms, J.} and {de Wolf, E.} and {Yatkin, K.} and {Yepes, H.} and {Zornoza, J. D.} and {Zúñiga, J.} and {Baerwald, P.}},
	title = {{Search for muon neutrinos from gamma-ray bursts with the ANTARES neutrino telescope using 2008 to 2011 data}},
	DOI= "10.1051/0004-6361/201322169",
	journal = {A\&A},
	year = 2013,
	volume = 559,
	pages = "A9",
}

@article{ANTICE,
      author         = "Albert, A. and Andr{\'e}, M. and Anghinolfi, M. and others",
      collaboration  = "ANTARES and IceCube",
      title          = "{ANTARES and IceCube Combined Search for Neutrino Point-like and Extended Sources in the Southern Sky}",
      journal        = "The Astrophysical Journal",
      volume         = "892",
      year           = "2020",
      number         = "2",
      pages          = "92",
      doi            = "10.3847/1538-4357/ab7afb",
      eprint         = "2001.06415",
      archivePrefix  = "arXiv",
      primaryClass   = "astro-ph.HE",
}

@article{ICE2016,
      author         = "Aartsen, M. G. and Abraham, K. and Ackermann, M. and others",
      collaboration  = "IceCube",
      title          = "{An All-sky Search for Three Flavors of Neutrinos from Gamma-ray Bursts with the IceCube Neutrino Observatory}",
      journal        = "The Astrophysical Journal",
      volume         = "824",
      year           = "2016",
      number         = "2",
      pages          = "115",
      doi            = "10.3847/0004-637X/824/2/115",
      eprint         = "1601.06441",
      archivePrefix  = "arXiv",
      primaryClass   = "astro-ph.HE",
}

@ARTICLE{ICE2022,
       author = {{Abbasi}, R. and {Ackermann}, M. and {Adams}, J. and {Aguilar}, J.~A. and {Ahlers}, M. and {Ahrens}, M. and {Alameddine}, J.~M. and {Alves}, A.~A. and {Amin}, N.~M. and {Andeen}, K. and {Anderson}, T. and {Anton}, G. and {Arg{\"u}elles}, C. and {Ashida}, Y. and {Athanasiadou}, S. and {Axani}, S. and {Bai}, X. and {Balagopal}, A.~V. and {Barwick}, S.~W. and {Basu}, V. and {Baur}, S. and {Bay}, R. and {Beatty}, J.~J. and {Becker}, K.-H. and {Becker Tjus}, J. and {Beise}, J. and {Bellenghi}, C. and {Benda}, S. and {BenZvi}, S. and {Berley}, D. and {Bernardini}, E. and {Besson}, D.~Z. and {Binder}, G. and {Bindig}, D. and {Blaufuss}, E. and {Blot}, S. and {Boddenberg}, M. and {Bontempo}, F. and {Book}, J.~Y. and {Borowka}, J. and {B{\"o}ser}, S. and {Botner}, O. and {B{\"o}ttcher}, J. and {Bourbeau}, E. and {Bradascio}, F. and {Braun}, J. and {Brinson}, B. and {Bron}, S. and {Brostean-Kaiser}, J. and {Burley}, R.~T. and {Busse}, R.~S. and {Campana}, M.~A. and {Carnie-Bronca}, E.~G. and {Chen}, C. and {Chen}, Z. and {Chirkin}, D. and {Choi}, K. and {Clark}, B.~A. and {Clark}, K. and {Classen}, L. and {Coleman}, A. and {Collin}, G.~H. and {Connolly}, A. and {Conrad}, J.~M. and {Coppin}, P. and {Correa}, P. and {Cowen}, D.~F. and {Cross}, R. and {Dappen}, C. and {Dave}, P. and {De Clercq}, C. and {DeLaunay}, J.~J. and {L{\'o}pez}, D. Delgado and {Dembinski}, H. and {Deoskar}, K. and {Desai}, A. and {Desiati}, P. and {de Vries}, K.~D. and {de Wasseige}, G. and {DeYoung}, T. and {Diaz}, A. and {D{\'\i}az-V{\'e}lez}, J.~C. and {Dittmer}, M. and {Dujmovic}, H. and {DuVernois}, M.~A. and {Ehrhardt}, T. and {Eller}, P. and {Engel}, R. and {Erpenbeck}, H. and {Evans}, J. and {Evenson}, P.~A. and {Fan}, K.~L. and {Fazely}, A.~R. and {Fedynitch}, A. and {Feigl}, N. and {Fiedlschuster}, S. and {Fienberg}, A.~T. and {Finley}, C. and {Fischer}, L. and {Fox}, D. and {Franckowiak}, A. and {Friedman}, E. and {Fritz}, A. and {F{\"u}rst}, P. and {Gaisser}, T.~K. and {Gallagher}, J. and {Ganster}, E. and {Garcia}, A. and {Garrappa}, S. and {Gerhardt}, L. and {Ghadimi}, A. and {Glaser}, C. and {Glauch}, T. and {Gl{\"u}senkamp}, T. and {Goehlke}, N. and {Gonzalez}, J.~G. and {Goswami}, S. and {Grant}, D. and {Gr{\'e}goire}, T. and {Griswold}, S. and {G{\"u}nther}, C. and {Gutjahr}, P. and {Haack}, C. and {Hallgren}, A. and {Halliday}, R. and {Halve}, L. and {Halzen}, F. and {Ha Minh}, M. and {Hanson}, K. and {Hardin}, J. and {Harnisch}, A.~A. and {Haungs}, A. and {Helbing}, K. and {Henningsen}, F. and {Hettinger}, E.~C. and {Hickford}, S. and {Hignight}, J. and {Hill}, C. and {Hill}, G.~C. and {Hoffman}, K.~D. and {Hoshina}, K. and {Hou}, W. and {Huber}, M. and {Huber}, T. and {Hultqvist}, K. and {H{\"u}nnefeld}, M. and {Hussain}, R. and {Hymon}, K. and {In}, S. and {Iovine}, N. and {Ishihara}, A. and {Jansson}, M. and {Japaridze}, G.~S. and {Jeong}, M. and {Jin}, M. and {Jones}, B.~J.~P. and {Kang}, D. and {Kang}, W. and {Kang}, X. and {Kappes}, A. and {Kappesser}, D. and {Kardum}, L. and {Karg}, T. and {Karl}, M. and {Karle}, A. and {Katz}, U. and {Kauer}, M. and {Kellermann}, M. and {Kelley}, J.~L. and {Kheirandish}, A. and {Kin}, K. and {Kiryluk}, J. and {Klein}, S.~R. and {Kochocki}, A. and {Koirala}, R. and {Kolanoski}, H. and {Kontrimas}, T. and {K{\"o}pke}, L. and {Kopper}, C. and {Kopper}, S. and {Koskinen}, D.~J. and {Koundal}, P. and {Kovacevich}, M. and {Kowalski}, M. and {Kozynets}, T. and {Krupczak}, E. and {Kun}, E. and {Kurahashi}, N. and {Lad}, N. and {Lagunas Gualda}, C. and {Larson}, M.~J. and {Lauber}, F. and {Lazar}, J.~P. and {Lee}, J.~W. and {Leonard}, K. and {Leszczy{\'n}ska}, A. and {Li}, Y. and {Lincetto}, M. and {Liu}, Q.~R. and {Liubarska}, M.},
        title = "{Searches for Neutrinos from Gamma-Ray Bursts Using the IceCube Neutrino Observatory}",
      journal = {ApJ},
         year = 2022,
       volume = {939},
       number = {2},
          eid = {116},
        pages = {116},
          doi = {10.3847/1538-4357/ac9785},
archivePrefix = {arXiv},
       eprint = {2205.11410},
 primaryClass = {astro-ph.HE},
}

@article{Aartsen2017,
  author = {Aartsen, M. G. and Abraham, K. and Ackermann, M. and Adams, J. and Aguilar, J. A. and Ahlers, M. and Ahrens, M. and Altmann, D. and Andeen, K. and Anderson, T. and Ansseau, I. and Anton, G. and Archinger, M. and Arg\"uelles, C. and Auffenberg, J. and Axani, S. and Bai, X. and Barwick, S. W. and Baum, V. and Bay, R. and Beatty, J. J. and Becker Tjus, J. and Becker, K.-H. and BenZvi, S. and Berley, D. and Bernardini, E. and Bernhard, A. and Besson, D. Z. and Binder, G. and Bindig, D. and Bissok, M. and Blaufuss, E. and Blot, S. and Bohm, C. and B\"orner, M. and Bos, F. and Bose, D. and B\"oser, S. and Botner, O. and Braun, J. and Brayeur, L. and Bretz, H.-P. and Bron, S. and Burgman, A. and Carver, T. and Casier, M. and Cheung, E. and Chirkin, D. and Christov, A. and Clark, K. and Classen, L. and Coenders, S. and Collin, G. H. and Conrad, J. M. and Cowen, D. F. and Cross, R. and Day, M. and de Andr\'e, J. P. A. M. and De Clercq, C. and del Pino Rosendo, E. and Dembinski, H. and De Ridder, S. and Desiati, P. and de Vries, K. D. and de Wasseige, G. and de With, M. and DeYoung, T. and D\'iaz-V\'elez, J. C. and di Lorenzo, V. and Dujmovic, H. and Dumm, J. P. and Dunkman, M. and Eberhardt, B. and Ehrhardt, T. and Eichmann, B. and Eller, P. and Euler, S. and Evenson, P. A. and Fahey, S. and Fazely, A. R. and Feintzeig, J. and Felde, J. and Filimonov, K. and Finley, C. and Flis, S. and F\"osig, C.-C. and Franckowiak, A. and Friedman, E. and Fuchs, T. and Gaisser, T. K. and Gallagher, J. and Gerhardt, L. and Ghorbani, K. and Giang, W. and Gladstone, L. and Glauch, T. and Gl\"usenkamp, T. and Goldschmidt, A. and Golup, G. and Gonzalez, J. G. and Grant, D. and Griffith, Z. and Haack, C. and Haj Ismail, A. and Hallgren, A. and Halzen, F. and Hansen, E. and Hansmann, T. and Hanson, K. and Hebecker, D. and Heereman, D. and Helbing, K. and Hellauer, R. and Hickford, S. and Hignight, J. and Hill, G. C. and Hoffman, K. D. and Hoffmann, R. and Holzapfel, K. and Hoshina, K. and Huang, F. and Huber, M. and Hultqvist, K. and In, S. and Ishihara, A. and Jacobi, E. and Japaridze, G. S. and Jeong, M. and Jero, K. and Jones, B. J. P. and Jurkovic, M. and Kappes, A. and Karg, T. and Karle, A. and Katz, U. and Kauer, M. and Keivani, A. and Kelley, J. L. and Kheirandish, A. and Kim, M. and Kintscher, T. and Kiryluk, J. and Kittler, T. and Klein, S. R. and Kohnen, G. and Koirala, R. and Kolanoski, H. and Konietz, R. and K\"opke, L. and Kopper, C. and Kopper, S. and Koskinen, D. J. and Kowalski, M. and Krings, K. and Kroll, M. and Kr\"uckl, G. and Kr\"uger, C. and Kunnen, J. and Kunwar, S. and Kurahashi, N. and Kuwabara, T. and Labare, M. and Lanfranchi, J. L. and Larson, M. J. and Lauber, F. and Lennarz, D. and Lesiak-Bzdak, M. and Leuermann, M. and Lu, L. and L\"unemann, J. and Madsen, J. and Maggi, G. and Mahn, K. B. M. and Mancina, S. and Mandelartz, M. and Maruyama, R. and Mase, K. and Maunu, R. and McNally, F. and Meagher, K. and Medici, M. and Meier, M. and Meli, A. and Menne, T. and Merino, G. and Meures, T. and Miarecki, S. and Mohrmann, L. and Montaruli, T. and Moulai, M. and Nahnhauer, R. and Naumann, U. and Neer, G. and Niederhausen, H. and Nowicki, S. C. and Nygren, D. R. and Obertacke Pollmann, A. and Olivas, A. and O'Murchadha, A. and Palczewski, T. and Pandya, H. and Pankova, D. V. and Peiffer, P. and Penek, \"O. and Pepper, J. A. and P\'erez de los Heros, C. and Pieloth, D. and Pinat, E. and Price, P. B. and Przybylski, G. T. and Quinnan, M. and Raab, C. and R\"adel, L. and Rameez, M. and Rawlins, K. and Reimann, R. and Relethford, B. and Relich, M. and Resconi, E. and Rhode, W. and Richman, M. and Riedel, B. and Robertson, S. and Rongen, M. and Rott, C. and Ruhe, T. and Ryckbosch, D. and Rysewyk, D. and Sabbatini, L. and Sanchez Herrera, S. E. and Sandrock, A. and Sandroos, J. and Sarkar, S. and Satalecka, K. and Schlunder, P. and Schmidt, T. and Schoenen, S. and Sch\"oneberg, S. and Schumacher, L. and Seckel, D. and Seunarine, S. and Soldin, D. and Song, M. and Spiczak, G. M. and Spiering, C. and Stanev, T. and Stasik, A. and Stettner, J. and Steuer, A. and Stezelberger, T. and Stokstad, R. G. and St\"ossl, A. and Str\"om, R. and Strotjohann, N. L. and Sullivan, G. W. and Sutherland, M. and Taavola, H. and Taboada, I. and Tatar, J. and Tenholt, F. and Ter-Antonyan, S. and Terliuk, A. and Te\v{s}i\'c, G. and Tilav, S. and Toale, P. A. and Tobin, M. N. and Toscano, S. and Tosi, D. and Tselengidou, M. and Turcati, A. and Unger, E. and Usner, M. and Vandenbroucke, J. and van Eijndhoven, N. and Vanheule, S. and van Rossem, M. and van Santen, J. and Veenkamp, J. and Vehring, M. and Voge, M. and Vogel, E. and Vraeghe, M. and Walck, C. and Wallace, A. and Wallraff, M. and Wandkowsky, N. and Weaver, Ch. and Weiss, M. J. and Wendt, C. and Westerhoff, S. and Whelan, B. J. and Wickmann, S. and Wiebe, K. and Wiebusch, C. H. and Wille, L. and Williams, D. R. and Wills, L. and Wolf, M. and Wood, T. R. and Woolsey, E. and Woschnagg, K. and Xu, D. L. and Xu, X. W. and Xu, Y. and Yanez, J. P. and Yodh, G. and Yoshida, S. and Zoll, M. and IceCube Collaboration},
  title = {{All-sky Search for Time-integrated Neutrino Emission from Astrophysical Sources with 7 yr of IceCube Data}},
  journal = {Astrophys. J.},
  volume = {835},
  number = {2},
  pages = {151},
  year = {2017},
  doi = {10.3847/1538-4357/835/2/151}
}

@article{Masudepjc,
  author  = {Brevik, Iver and Chaichian, Masud and Oksanen, Markku},
  title   = {{Dispersion of light traveling through the interstellar space, induced and intrinsic Lorentz invariance violation}},
  journal = {Eur. Phys. J. C},
  year    = {2021},
  volume  = {81},
  pages   = {926},
  doi     = {10.1140/epjc/s10052-021-09707-3}
}

@article{Masudplb,
  author  = {Brevik, Iver H. and Chaichian, Moshe M. and Tureanu, Anca},
  title   = {{Below the Schwinger critical magnetic field value, quantum vacuum and gamma-ray bursts delay}},
  journal = {Phys. Lett. B},
  year    = {2025},
  volume  = {861},
  pages   = {139272},
  doi     = {10.1016/j.physletb.2025.139272}
}

@article{Nat.Astro,
  author    = {Zhang, S.-N. and Kole, M. and Bao, T.-W. and others},
  title     = {{Detailed polarization measurements of the prompt emission of five gamma-ray bursts}},
  journal   = {Nature Astronomy},
  volume    = {3},
  pages     = {258--264},
  year      = {2019},
  doi       = {10.1038/s41550-018-0664-0},
}

@article{Chattopadhyay_2019,
doi = {10.3847/1538-4357/ab40b7},
year = {2019},
publisher = {The American Astronomical Society},
volume = {884},
number = {2},
pages = {123},
author = {Chattopadhyay, Tanmoy and Vadawale, Santosh V. and Aarthy, E. and Mithun, N. P. S. and Chand, Vikas and Ratheesh, Ajay and Basak, Rupal and Rao, A. R. and Bhalerao, Varun and Mate, Sujay and B., Arvind and Sharma, V. and Bhattacharya, Dipankar},
title = {{Prompt Emission Polarimetry of Gamma-Ray Bursts with the AstroSat CZT Imager}},
journal = {The Astrophysical Journal},
}

@article{JCAP2016,
doi = {10.1088/1475-7516/2016/08/031},
year = {2016},
volume = {2016},
number = {08},
pages = {031},
author = {Wei, Jun-Jie and Wu, Xue-Feng and Gao, He and Mészáros, Peter},
title = {{Limits on the neutrino velocity, Lorentz invariance, and the weak equivalence principle with TeV neutrinos from gamma-ray bursts}},
journal = {Journal of Cosmology and Astroparticle Physics},
}

@article{Catalog_2020,
    doi= {10.3847/1538-4357/ab7a18},
	year = 2020,
	publisher = {The American Astronomical Society},
	volume = {893},
	number = {1},
	pages = {46},
	author = {A. von Kienlin and C. A. Meegan and W. S. Paciesas and P. N. Bhat and E. Bissaldi and M. S. Briggs and E. Burns and W. H. Cleveland and M. H. Gibby and M. M. Giles and A. Goldstein and R. Hamburg and C. M. Hui and D. Kocevski and B. Mailyan and C. Malacaria and S. Poolakkil and R. D. Preece and O. J. Roberts and P. Veres and C. A. Wilson-Hodge},
	title = {{The Fourth {Fermi-GBM} Gamma-Ray Burst Catalog: A Decade of Data}},
	journal = {The Astrophysical Journal}
}

@article{DESI2024,
  author    = {A. G. Adame and J. Aguilar and S. Ahlen and S. Alam and D. M. Alexander and M. Alvarez and O. Alves and A. Anand and U. Andrade and E. Armengaud},
  title     = {{DESI 2024 VI: cosmological constraints from the measurements of baryon acoustic oscillations}},
  journal   = {JCAP},
  year      = {2025},
  volume    = {2025},
  number    = {02},
  pages     = {021},
  doi       = {10.1088/1475-7516/2025/02/021},
  publisher = {IOP Publishing}
}

@article{Planck2018,
  author       = {{Planck Collaboration} and {Aghanim, N.} and {Akrami, Y.} and {Ashdown, M.} and {Aumont, J.} and {Baccigalupi, C.} and {Ballardini, M.} and {Banday, A. J.} and {Barreiro, R. B.} and {Bartolo, N.} and {Basak, S.} and {Battye, R.} and {Benabed, K.} and {Bernard, J.-P.} and {Bersanelli, M.} and {Bielewicz, P.} and {Bock, J. J.} and {Bond, J. R.} and {Borrill, J.} and {Bouchet, F. R.} and {Boulanger, F.} and {Bucher, M.} and {Burigana, C.} and {Butler, R. C.} and {Calabrese, E.} and {Cardoso, J.-F.} and {Carron, J.} and {Challinor, A.} and {Chiang, H. C.} and {Chluba, J.} and {Colombo, L. P. L.} and {Combet, C.} and {Contreras, D.} and {Crill, B. P.} and {Cuttaia, F.} and {de Bernardis, P.} and {de Zotti, G.} and {Delabrouille, J.} and {Delouis, J.-M.} and {Di Valentino, E.} and {Diego, J. M.} and {Doré, O.} and {Douspis, M.} and {Ducout, A.} and {Dupac, X.} and {Dusini, S.} and {Efstathiou, G.} and {Elsner, F.} and {Enßlin, T. A.} and {Eriksen, H. K.} and {Fantaye, Y.} and {Farhang, M.} and {Fergusson, J.} and {Fernandez-Cobos, R.} and {Finelli, F.} and {Forastieri, F.} and {Frailis, M.} and {Fraisse, A. A.} and {Franceschi, E.} and {Frolov, A.} and {Galeotta, S.} and {Galli, S.} and {Ganga, K.} and {Génova-Santos, R. T.} and {Gerbino, M.} and {Ghosh, T.} and {González-Nuevo, J.} and {Górski, K. M.} and {Gratton, S.} and {Gruppuso, A.} and {Gudmundsson, J. E.} and {Hamann, J.} and {Handley, W.} and {Hansen, F. K.} and {Herranz, D.} and {Hildebrandt, S. R.} and {Hivon, E.} and {Huang, Z.} and {Jaffe, A. H.} and {Jones, W. C.} and {Karakci, A.} and {Keihänen, E.} and {Keskitalo, R.} and {Kiiveri, K.} and {Kim, J.} and {Kisner, T. S.} and {Knox, L.} and {Krachmalnicoff, N.} and {Kunz, M.} and {Kurki-Suonio, H.} and {Lagache, G.} and {Lamarre, J.-M.} and {Lasenby, A.} and {Lattanzi, M.} and {Lawrence, C. R.} and {Le Jeune, M.} and {Lemos, P.} and {Lesgourgues, J.} and {Levrier, F.} and {Lewis, A.} and {Liguori, M.} and {Lilje, P. B.} and {Lilley, M.} and {Lindholm, V.} and {López-Caniego, M.} and {Lubin, P. M.} and {Ma, Y.-Z.} and {Macías-Pérez, J. F.} and {Maggio, G.} and {Maino, D.} and {Mandolesi, N.} and {Mangilli, A.} and {Marcos-Caballero, A.} and {Maris, M.} and {Martin, P. G.} and {Martinelli, M.} and {Martínez-González, E.} and {Matarrese, S.} and {Mauri, N.} and {McEwen, J. D.} and {Meinhold, P. R.} and {Melchiorri, A.} and {Mennella, A.} and {Migliaccio, M.} and {Millea, M.} and {Mitra, S.} and {Miville-Deschênes, M.-A.} and {Molinari, D.} and {Montier, L.} and {Morgante, G.} and {Moss, A.} and {Natoli, P.} and {Nørgaard-Nielsen, H. U.} and {Pagano, L.} and {Paoletti, D.} and {Partridge, B.} and {Patanchon, G.} and {Peiris, H. V.} and {Perrotta, F.} and {Pettorino, V.} and {Piacentini, F.} and {Polastri, L.} and {Polenta, G.} and {Puget, J.-L.} and {Rachen, J. P.} and {Reinecke, M.} and {Remazeilles, M.} and {Renzi, A.} and {Rocha, G.} and {Rosset, C.} and {Roudier, G.} and {Rubiño-Martín, J. A.} and {Ruiz-Granados, B.} and {Salvati, L.} and {Sandri, M.} and {Savelainen, M.} and {Scott, D.} and {Shellard, E. P. S.} and {Sirignano, C.} and {Sirri, G.} and {Spencer, L. D.} and {Sunyaev, R.} and {Suur-Uski, A.-S.} and {Tauber, J. A.} and {Tavagnacco, D.} and {Tenti, M.} and {Toffolatti, L.} and {Tomasi, M.} and {Trombetti, T.} and {Valenziano, L.} and {Valiviita, J.} and {Van Tent, B.} and {Vibert, L.} and {Vielva, P.} and {Villa, F.} and {Vittorio, N.} and {Wandelt, B. D.} and {Wehus, I. K.} and {White, M.} and {White, S. D. M.} and {Zacchei, A.} and {Zonca, A.}},
  title        = {Planck 2018 results. {VI}. {C}osmological parameters},
  journal      = {Astron. Astrophys.},
  volume       = {641},
  pages        = {A6},
  year         = {2020},
  eprint       = {1807.06209},
  archivePrefix= {arXiv},
  primaryClass = {astro-ph.CO},
  doi          = {10.1051/0004-6361/201833910}
}

@article{Helayel2,
  author  = {Paix{\~a}o, J. M. A. and Ospedal, L. P. R. and Neves, M. J. and Helay{\"e}l-Neto, J. A.},
  title   = {{Probing the interference between non-linear, axionic and space-time-anisotropy effects in the QED vacuum}},
  journal = {JHEP},
  year    = {2024},
  volume  = {05},
  pages   = {029},
  doi     = {10.1007/JHEP05(2024)029}
}

@article{Helayel1,
  author  = {Paix{\~a}o, J. M. A. and Ospedal, L. P. R. and Neves, M. J. and Helay{\"e}l-Neto, J. A.},
  title   = {The axion-photon mixing in non-linear electrodynamic scenarios},
  journal = {JHEP},
  year    = {2022},
  volume  = {10},
  pages   = {160},
  doi     = {10.1007/JHEP10(2022)160}
}

@article{Ouellet2019,
  title = {Solutions to axion electrodynamics in various geometries},
  author = {Ouellet, Jonathan and Bogorad, Zachary},
  journal = {Phys. Rev. D},
  volume = {99},
  issue = {5},
  pages = {055010},
  year = {2019},
  publisher = {American Physical Society},
  doi = {10.1103/PhysRevD.99.055010},
}

@article{Optical,
  title = {Optical properties of dynamical axion backgrounds},
  author = {McDonald, Jamie I. and Ventura, Lu\'{\i}s B.},
  journal = {Phys. Rev. D},
  volume = {101},
  issue = {12},
  pages = {123503},
  year = {2020},
  publisher = {American Physical Society},
  doi = {10.1103/PhysRevD.101.123503},
}

@article{Mixing1988,
  title = {Mixing of the photon with low-mass particles},
  author = {Raffelt, Georg and Stodolsky, Leo},
  journal = {Phys. Rev. D},
  volume = {37},
  issue = {5},
  pages = {1237--1249},
  numpages = {0},
  year = {1988},
  month = {Mar},
  publisher = {American Physical Society},
  doi = {10.1103/PhysRevD.37.1237},
  url = {https://link.aps.org/doi/10.1103/PhysRevD.37.1237}
}

@article{Catena2009,
  author  = {Catena, R. and Ullio, P.},
  title   = {A novel determination of the local dark matter density},
  journal = {JCAP},
  year    = {2010},
  volume  = {08},
  pages   = {004},
  doi     = {10.1088/1475-7516/2010/08/004}
}

@article{Sofue2020,
  author  = {Sofue, Yoshiaki},
  title   = {{Rotation Curve of the Milky Way and the Dark Matter Density}},
  journal = {Galaxies},
  year    = {2020},
  volume  = {8},
  number  = {2},
  pages   = {37},
  doi     = {10.3390/galaxies8020037},
}

@article{Iocco2011,
  author  = {Iocco, Fabio and Pato, Miguel and Bertone, Gianfranco and Jetzer, Philippe},
  title   = {{Dark Matter distribution in the Milky Way: microlensing and dynamical constraints}},
  journal = {JCAP},
  year    = {2011},
  volume  = {11},
  pages   = {029},
  doi     = {10.1088/1475-7516/2011/11/029}
}

@article{Irastorza2018,
  author  = {Irastorza, Igor G. and Redondo, Javier},
  title   = {New experimental approaches in the search for axion-like particles},
  journal = {Progress in Particle and Nuclear Physics},
  volume  = {102},
  pages   = {89-159},
  year    = {2018},
  doi     = {10.1016/j.ppnp.2018.05.003},
  eprint  = {1801.08127},
  archivePrefix = {arXiv},
  primaryClass = {hep-ph}
}

@article{Payez2015,
  author  = {Payez, Alexandre and Evoli, Carmelo and Fischer, Tobias and Giannotti, Maurizio and Mirizzi, Alessandro and Ringwald, Andreas},
  title   = {{Revisiting the SN1987A gamma-ray limit on ultralight axion-like particles}},
  journal = {JCAP},
  year    = {2015},
  volume  = {02},
  pages   = {006},
  doi     = {10.1088/1475-7516/2015/02/006},
  eprint  = {1410.3747},
  archivePrefix = {arXiv},
  primaryClass = {astro-ph.HE}
}

@article{AmelinoCamelia,
  author  = {Amelino-Camelia, Giovanni and Ellis, John and Mavromatos, N. E. and Nanopoulos, D. V. and Sarkar, Subir},
  title   = {Tests of quantum gravity from observations of gamma-ray bursts},
  journal = {Nature},
  year    = {1998},
  volume  = {393},
  number  = {6687},
  pages   = {763--765},
  doi     = {10.1038/31647}
}

@ARTICLE{Mavromatos20105409,
	author = {Mavromatos, Nick E.},
	title = {String quantum gravity, lorentz-invariance violation and gamma ray astronomy},
	year = {2010},
	journal = {International Journal of Modern Physics A},
	volume = {25},
	number = {30},
	pages = {5409 – 5485},
	doi = {10.1142/S0217751X10050792},
	
}

@article{RodriguezMartinez2006,
  author  = {Rodr{\'i}guez Mart{\'i}nez, Mar{\'i}a and Piran, Tsvi},
  title   = {{Constraining Lorentz violations with gamma ray bursts}},
  journal = {JCAP},
  year    = {2006},
  volume  = {04},
  pages   = {006},
  doi     = {10.1088/1475-7516/2006/04/006},
}

@article{IceCubeLIV,
  author  = {{IceCube Collaboration}},
  title   = {{Neutrino interferometry for high-precision tests of Lorentz symmetry with IceCube}},
  journal = {Nature Physics},
  volume  = {14},
  pages   = {961--966},
  year    = {2018},
  doi     = {10.1038/s41567-018-0172-2}
}

@article{Song2024,
  author  = {Song, Hanlin and Ma, Bo-Qiang},
  title   = {{Energy-dependent intrinsic time delay of gamma-ray bursts on testing Lorentz invariance violation}},
  journal = {Physics Letters B},
  volume  = {856},
  pages   = {138951},
  year    = {2024},
  doi     = {10.1016/j.physletb.2024.138951}
}

@ARTICLE{Cao2024,
	author = {Cao, Zhen and Aharonian, F. and Axikegu and Bai, Y.X. and Bao, Y.W. and Bastieri, D. and Bi, X.J. and Bi, Y.J. and Bian, W. and Bukevich, A.V. and Cao, Q. and Cao, W.Y. and Cao, Zhe and Chang, J. and Chang, J.F. and Chen, A.M. and Chen, E.S. and Chen, H.X. and Chen, Liang and Chen, Lin and Chen, Long and Chen, M.J. and Chen, M.L. and Chen, Q.H. and Chen, S. and Chen, S.H. and Chen, S.Z. and Chen, T.L. and Chen, Y. and Cheng, N. and Cheng, Y.D. and Cui, M.Y. and Cui, S.W. and Cui, X.H. and Cui, Y.D. and Dai, B.Z. and Dai, H.L. and Dai, Z.G. and Danzengluobu and Dong, X.Q. and Duan, K.K. and Fan, J.H. and Fan, Y.Z. and Fang, J. and Fang, J.H. and Fang, K. and Feng, C.F. and Feng, H. and Feng, L. and Feng, S.H. and Feng, X.T. and Feng, Y. and Feng, Y.L. and Gabici, S. and Gao, B. and Gao, C.D. and Gao, Q. and Gao, W. and Ge, M.M. and Geng, L.S. and Giacinti, G. and Gong, G.H. and Gou, Q.B. and Gu, M.H. and Guo, F.L. and Guo, X.L. and Guo, Y.Q. and Guo, Y.Y. and Han, Y.A. and Hasan, M. and He, H.H. and He, H.N. and He, J.Y. and He, Y. and Hor, Y.K. and Hou, B.W. and Hou, C. and Hou, X. and Hu, H.B. and Hu, Q. and Hu, S.C. and Huang, D.H. and Huang, T.Q. and Huang, W.J. and Huang, X.T. and Huang, X.Y. and Huang, Y. and Ji, X.L. and Jia, H.Y. and Jia, K. and Jiang, K. and Jiang, X.W. and Jiang, Z.J. and Jin, M. and Kang, M.M. and Karpikov, I. and Kuleshov, D. and Kurinov, K. and Li, B.B. and Li, C.M. and Li, Cheng and Li, Cong and Li, D. and Li, F. and Li, H.B. and Li, H.C. and Li, Jian and Li, Jie and Li, K. and Li, S.D. and Li, W.L. and Li, W.L. and Li, X.R. and Li, Xin and Li, Y.Z. and Li, Zhe and Li, Zhuo and Liang, E.W. and Liang, Y.F. and Lin, S.J. and Liu, B. and Liu, C. and Liu, D. and Liu, D.B. and Liu, H. and Liu, H.D. and Liu, J. and Liu, M.Y. and Liu, R.Y. and Liu, S.M. and Liu, W. and Liu, Y. and Liu, Y.N. and Luo, Q. and Luo, Y. and Lv, H.K. and Ma, B.Q. and Ma, L.L. and Ma, X.H. and Mao, J.R. and Min, Z. and Mitthumsiri, W. and Mu, H.J. and Nan, Y.C. and Neronov, A. and Ou, L.J. and Pattarakijwanich, P. and Pei, Z.Y. and Qi, J.C. and Qi, M.Y. and Qiao, B.Q. and Qin, J.J. and Raza, A. and Ruffolo, D. and Sáiz, A. and Saeed, M. and Semikoz, D. and Shao, L. and Shchegolev, O. and Sheng, X.D. and Shu, F.W. and Song, H.C. and Stenkin, Yu. V. and Stepanov, V. and Su, Y. and Sun, D.X. and Sun, Q.N. and Sun, X.N. and Sun, Z.B. and Takata, J. and Tam, P.H.T. and Tang, Q.W. and Tang, R. and Tang, Z.B. and Tian, W.W. and Wang, C. and Wang, C.B. and Wang, G.W. and Wang, H.G. and Wang, H.H. and Wang, J.C. and Wang, Kai and Wang, L.P. and Wang, L.Y. and Wang, P.H. and Wang, R. and Wang, W. and Wang, X.G. and Wang, X.Y. and Wang, Y. and Wang, Y.D. and Wang, Y.J. and Wang, Z.H. and Wang, Z.X. and Wang, Zhen and Wang, Zheng and Wei, D.M. and Wei, J.J. and Wei, Y.J. and Wen, T. and Wu, C.Y. and Wu, H.R. and Wu, Q.W. and Wu, S. and Wu, X.F. and Wu, Y.S. and Xi, S.Q. and Xia, J. and Xiang, G.M. and Xiao, D.X. and Xiao, G. and Xin, Y.L. and Xing, Y. and Xiong, D.R. and Xiong, Z. and Xu, D.L. and Xu, R.F. and Xu, R.X. and Xu, W.L. and Xue, L. and Yan, D.H. and Yan, J.Z. and Yan, T. and Yang, C.W. and Yang, C.Y. and Yang, F. and Yang, F.F. and Yang, L.L. and Yang, M.J. and Yang, R.Z. and Yang, W.X. and Yao, Y.H. and Yao, Z.G. and Yin, L.Q. and Yin, N. and You, X.H. and You, Z.Y. and Yu, Y.H. and Yuan, Q. and Yue, H. and Zeng, H.D. and Zeng, T.X. and Zeng, W. and Zha, M. and Zhang, B.B. and Zhang, F. and Zhang, H. and Zhang, H.M. and Zhang, H.Y. and Zhang, J.L. and Zhang, Li and Zhang, P.F. and Zhang, P.P. and Zhang, R. and Zhang, S.B. and Zhang, S.R. and Zhang, S.S. and Zhang, X. and Zhang, X.P. and Zhang, Y.F. and Zhang, Yi and Zhang, Yong and Zhao, B. and Zhao, J. and Zhao, L. and Zhao, L.Z. and Zhao, S.P. and Zhao, X.H. and Zheng, F. and Zhong, W.J. and Zhou, B. and Zhou, H. and Zhou, J.N. and Zhou, M. and Zhou, P. and Zhou, R. and Zhou, X.X. and Zhou, X.X. and Zhu, B.Y. and Zhu, C.G. and Zhu, F.R. and Zhu, H. and Zhu, K.J. and Zou, Y.C. and Zuo, X.},
	title = {{Stringent Tests of Lorentz Invariance Violation from LHAASO Observations of GRB 221009A}},
	year = {2024},
	journal = {Physical Review Letters},
	volume = {133},
    pages = {071501},
	number = {7},
	doi = {10.1103/PhysRevLett.133.071501},
}

@article{MADMAX,
  title = {{First Search for Axion Dark Matter with a MADMAX Prototype}},
  author = {Ary dos Santos Garcia, B. and Bergermann, D. and Caldwell, A. and Dabhi, V. and Diaconu, C. and Diehl, J. and Dvali, G. and Egge, J. and Garutti, E. and Heyminck, S. and Hubaut, F. and Ivanov, A. and Jochum, J. and Knirck, S. and Kramer, M. and Kreikemeyer-Lorenzo, D. and Krieger, C. and Lee, C. and Leppla-Weber, D. and Li, X. and Lindner, A. and Majorovits, B. and Maldonado, J. P. A. and Martini, A. and Miyazaki, A. and \"Oz, E. and Pralavorio, P. and Raffelt, G. and Redondo, J. and Ringwald, A. and Schaffran, J. and Schmidt, A. and Steffen, F. and Strandhagen, C. and Usherov, I. and Wang, H. and Wieching, G.},
  collaboration = {{MADMAX}},
  journal = {Phys. Rev. Lett.},
  volume = {135},
  issue = {4},
  pages = {041001},
  numpages = {7},
  year = {2025},
  month = {Jul},
  publisher = {American Physical Society},
  doi = {10.1103/c749-419q},
}

@article{IAXO2019,
  author  = {Armengaud, E. and Atti{\'e}, D. and Basso, S. and Brun, P. and Bykovskiy, N. and Carmona, J. M. and Castel, J. F. and Cebri{\'a}n, S. and Cicoli, M. and Civitani, M. and others},
  title   = {{Physics potential of the International Axion Observatory (IAXO)}},
  journal = {JCAP},
  year    = {2019},
  volume  = {06},
  pages   = {047},
  doi     = {10.1088/1475-7516/2019/06/047},
  eprint  = {1904.09155},
  archivePrefix = {arXiv},
  primaryClass = {physics.ins-det}
}

@article{Carenza2025,
  author  = {Carenza, P. and Garc{\'i}a Pascual, J. A. and Giannotti, M. and Irastorza, I. G. and Kaltschmidt, M. and Lella, A. and Lindner, A. and Lucente, G. and Mirizzi, A. and Puyuelo, M. J.},
  title   = {{Detecting supernova axions with IAXO}},
  journal = {JCAP},
  year    = {2025},
  volume  = {07},
  pages   = {075},
  doi     = {10.1088/1475-7516/2025/07/075}
}

@article{Mittal2025,
    author = "Mittal, Saurabh and Siegert, Thomas and Calore, Francesca and Carenza, Pierluca and Eisenberger, Laura and Giannotti, Maurizio and Lella, Alessandro and Mirizzi, Alessandro and Tsatsis, Dimitris and Yoneda, Hiroki",
    title = "{{Search for Axion-Like Particles from Nearby Pre-Supernova Stars}}",
    eprint = "2512.19298",
    archivePrefix = "arXiv",
    primaryClass = "astro-ph.HE",
    doi = "10.1051/0004-6361/202556951",
    journal = "Astron. Astrophys.",
    volume = "707",
    pages = "A108",
    year = "2026"}

@article{pulsar_polarimetry,
  title = {Searching for axions with time resolved pulsar polarimetry},
  author = {Chadha-Day, Francesca and Poddar, Tanmay Kumar},
  journal = {Phys. Rev. D},
  volume = {113},
  issue = {12},
  pages = {123014},
  numpages = {9},
  year = {2026},
  publisher = {APS},
  doi = {10.1103/vml6-f38z},
}

@article{Schwinger,
  title = {{On Gauge Invariance and Vacuum Polarization}},
  author = {Schwinger, Julian},
  journal = {Phys. Rev.},
  volume = {82},
  issue = {5},
  pages = {664--679},
  numpages = {0},
  year = {1951},
  publisher = {American Physical Society},
  doi = {10.1103/PhysRev.82.664},

}

@article{EH1936,
  author = {Heisenberg, Werner and Euler, Hans},
  title = {{Folgerungen aus der Diracschen Theorie des Positrons}},
  journal = {Z. Phys.},
  volume = {98},
  pages = {714-732},
  year = {1936},
  doi = {10.1007/BF01343663}
}

@misc{AxionLimits,
  author       = {Ciaran O'Hare},
  title        = {{cajohare/AxionLimits: AxionLimits}},
  year         = 2020,
  publisher    = {Zenodo},
  version      = {v1.0},
  doi          = {10.5281/zenodo.3932430},
  howpublished = {\url{https://cajohare.github.io/AxionLimits/}}
}

@article{Fiorillo:2026sn,
    author = "Fiorillo, Damiano F. G. and Gil Muyor, \'Angel and Janka, Hans-Thomas and Raffelt, Georg G. and Vitagliano, Edoardo",
    title = "{{Axion-photon conversion in transient compact stars: Systematics, constraints, and opportunities}}",
    journal = "JCAP",
    volume = "03",
    pages = "053",
    year = "2026",
    doi = "10.1088/1475-7516/2026/03/053",
    eprint = "2509.13322",
    archivePrefix = "arXiv",
    primaryClass = "hep-ph"
}

@article{Lawson:2019plasma,
  title = {Tunable Axion Plasma Haloscopes},
  author = {Lawson, Matthew and Millar, Alexander J. and Pancaldi, Matteo and Vitagliano, Edoardo and Wilczek, Frank},
  journal = {Phys. Rev. Lett.},
  volume = {123},
  issue = {14},
  pages = {141802},
  numpages = {7},
  year = {2019},
  publisher = {American Physical Society},
  doi = {10.1103/PhysRevLett.123.141802},
}

@article{Baring1998Radio-Quiet,
title={{Radio-Quiet Pulsars with Ultrastrong Magnetic Fields}},
author={M. Baring and A. Harding},
journal={The Astrophysical Journal Letters},
year={1998},
volume={507},
pages={L55 - L58},
doi={10.1086/311679}
}

@article{Vranesevic_2004,
doi = {10.1086/427208},
year = {2004},
month = {nov},
volume = {617},
number = {2},
pages = {L139},
author = {Vranesevic, N. and Manchester, R. N. and Lorimer, D. R. and Hobbs, G. B. and Lyne, A. G. and Kramer, M. and Camilo, F. and Stairs, I. H. and Kaspi, V. M. and D’Amico, N. and Possenti, A. and Crawford, F. and Faulkner, A. J. and McLaughlin, M. A.},
title = {{Pulsar Birthrates from the Parkes Multibeam Survey}},
journal = {The Astrophysical Journal},

}

@article{Xue_2023,
doi = {10.1088/1674-4527/acdbbd},
year = {2023},
month = {jul},
publisher = {National Astromonical Observatories, CAS and IOP Publishing},
volume = {23},
number = {9},
pages = {095005},
author = {Xue, Mengyao and Zhu, Weiwei and Wu, Xiangping and Xu, Renxin and Wang, Hongguang},
title = {{Pulsar Discovery Prospect of FASTA}},
journal = {Research in Astronomy and Astrophysics},

}

@article{AMEGO,
author = {Regina Caputo and Marco Ajello and Carolyn A. Kierans and Jeremy S. Perkins and Judith L. Racusin and Luca Baldini and Matthew G. Baring and Elisabetta Bissaldi and Eric Burns and Nicholas Cannady and Eric Charles and Rui M. Curado da Silva and Ke Fang and Henrike Fleischhack and Chris Fryer and Yasushi Fukazawa and J. Eric Grove and Dieter Hartmann and Eric J. Howell and Manoj Jadhav and Christopher M. Karwin and Daniel Kocevski and Naoko Kurahashi and Luca Latronico and Tiffany R. Lewis and Richard Leys and Amy Lien and Lea Marcotulli and Israel Martinez-Castellanos and Mario Nicola Mazziotta and Julie McEnery and Jessica Metcalfe and Kohta Murase and Michela Negro and Lucas Parker and Bernard Phlips and Chanda Prescod-Weinstein and Soebur Razzaque and Peter S. Shawhan and Yong Sheng and Tom A. Shutt and Daniel Shy and Clio Sleator and Amanda L. Steinhebel and Nicolas Striebig and Yusuke Suda and Donggeun Tak and Hiro Tajima and Janeth Valverde and Tonia M. Venters and Zorawar Wadiasingh and Richard S. Woolf and Eric A. Wulf and Haocheng Zhang and Andreas Zoglauer},
title = {{All-sky Medium Energy Gamma-ray Observatory eXplorer mission concept}},
volume = {8},
journal = {Journal of Astronomical Telescopes, Instruments, and Systems},
number = {4},
publisher = {SPIE},
pages = {044003},
year = {2022},
doi = {10.1117/1.JATIS.8.4.044003},
}

@article{Harari1992,
title = {{Effects of a Nambu-Goldstone boson on the polarization of radio galaxies and the cosmic microwave background}},
journal = {Physics Letters B},
volume = {289},
number = {1},
pages = {67-72},
year = {1992},
issn = {0370-2693},
doi = {https://doi.org/10.1016/0370-2693(92)91363-E},
author = {Diego Harari and Pierre Sikivie},
}

@article{Carroll1990,
  title = {{Limits on a Lorentz- and parity-violating modification of electrodynamics}},
  author = {Carroll, Sean M. and Field, George B. and Jackiw, Roman},
  journal = {Phys. Rev. D},
  volume = {41},
  issue = {4},
  pages = {1231--1240},
  numpages = {0},
  year = {1990},
  month = {Feb},
  publisher = {APS},
  doi = {10.1103/PhysRevD.41.1231},
}

@article{Fedderke2019,
  title = {{Axion dark matter detection with CMB polarization}},
  author = {Fedderke, Michael A. and Graham, Peter W. and Rajendran, Surjeet},
  journal = {Phys. Rev. D},
  volume = {100},
  issue = {1},
  pages = {015040},
  numpages = {24},
  year = {2019},
  publisher = {American Physical Society},
  doi = {10.1103/PhysRevD.100.015040},
}

@article{Lattimer2001,
  title={{Neutron Star Structure and the Equation of State}},
  author={Lattimer, James M and Prakash, Madappa},
  journal={The Astrophysical Journal},
  volume={550},
  number={1},
  pages={426},
  year={2001},
  publisher={IOP Publishing},
  doi={10.1086/319702}
}

@article{Berger2014,
  title={{Short-Duration Gamma-Ray Bursts}},
  author={Berger, Edo},
  journal={Annual Review of Astronomy and Astrophysics},
  volume={52},
  pages={43--105},
  year={2014},
  publisher={Annual Reviews},
  doi={https://doi.org/10.1146/annurev-astro-081913-035926}
}

@article{Ayala2014,
  title = {{Revisiting the Bound on Axion-Photon Coupling from Globular Clusters}},
  author = {Ayala, Adrian and Dom\'{\i}nguez, Inma and Giannotti, Maurizio and Mirizzi, Alessandro and Straniero, Oscar},
  journal = {Phys. Rev. Lett.},
  volume = {113},
  issue = {19},
  pages = {191302},
  numpages = {5},
  year = {2014},
  publisher = {APS},
  doi = {10.1103/PhysRevLett.113.191302},
}

@article{eASTROGAM2017,
title = {{Science with e-ASTROGAM: A space mission for MeV–GeV gamma-ray astrophysics}},
journal = {Journal of High Energy Astrophysics},
volume = {19},
pages = {1-106},
year = {2018},
issn = {2214-4048},
doi = {https://doi.org/10.1016/j.jheap.2018.07.001},
author = {A. {De Angelis} and V. Tatischeff and I.A. Grenier and J. McEnery and M. Mallamaci and M. Tavani and U. Oberlack and L. Hanlon and R. Walter and A. Argan and P. {Von Ballmoos} and A. Bulgarelli and A. Bykov and M. Hernanz and G. Kanbach and I. Kuvvetli and M. Pearce and A. Zdziarski and J. Conrad and G. Ghisellini and A. Harding and J. Isern and M. Leising and F. Longo and G. Madejski and M. Martinez and M.N. Mazziotta and J.M. Paredes and M. Pohl and R. Rando and M. Razzano and A. Aboudan and M. Ackermann and A. Addazi and M. Ajello and C. Albertus and J.M. Álvarez and G. Ambrosi and S. Antón and L.A. Antonelli and A. Babic and B. Baibussinov and M. Balbo and L. Baldini and S. Balman and C. Bambi and U. {Barres de Almeida} and J.A. Barrio and R. Bartels and D. Bastieri and W. Bednarek and D. Bernard and E. Bernardini and T. Bernasconi and B. Bertucci and A. Biland and E. Bissaldi and M. Boettcher and V. Bonvicini and V. Bosch-Ramon and E. Bottacini and V. Bozhilov and T. Bretz and M. Branchesi and V. Brdar and T. Bringmann and A. Brogna and C. {Budtz Jørgensen} and G. Busetto and S. Buson and M. Busso and A. Caccianiga and S. Camera and R. Campana and P. Caraveo and M. Cardillo and P. Carlson and S. Celestin and M. Cermeño and A. Chen and C.C. Cheung and E. Churazov and S. Ciprini and A. Coc and S. Colafrancesco and A. Coleiro and W. Collmar and P. Coppi and R. {Curado da Silva} and S. Cutini and F. D'Ammando and B. {De Lotto} and D. {de Martino} and A. {De Rosa} and M. {Del Santo} and L. Delgado and R. Diehl and S. Dietrich and A.D. Dolgov and A. Domínguez and D. {Dominis Prester} and I. Donnarumma and D. Dorner and M. Doro and M. Dutra and D. Elsaesser and M. Fabrizio and A. Fernández-Barral and V. Fioretti and L. Foffano and V. Formato and N. Fornengo and L. Foschini and A. Franceschini and A. Franckowiak and S. Funk and F. Fuschino and D. Gaggero and G. Galanti and F. Gargano and D. Gasparrini and R. Gehrz and P. Giammaria and N. Giglietto and P. Giommi and F. Giordano and M. Giroletti and G. Ghirlanda and N. Godinovic and C. Gouiffés and J.E. Grove and C. Hamadache and D.H. Hartmann and M. Hayashida and A. Hryczuk and P. Jean and T. Johnson and J. José and S. Kaufmann and B. Khelifi and J. Kiener and J. Knödlseder and M. Kole and J. Kopp and V. Kozhuharov and C. Labanti and S. Lalkovski and P. Laurent and O. Limousin and M. Linares and E. Lindfors and M. Lindner and J. Liu and S. Lombardi and F. Loparco and R. López-Coto and M. {López Moya} and B. Lott and P. Lubrano and D. Malyshev and N. Mankuzhiyil and K. Mannheim and M.J. Marchã and A. Marcianò and B. Marcote and M. Mariotti and M. Marisaldi and S. McBreen and S. Mereghetti and A. Merle and R. Mignani and G. Minervini and A. Moiseev and A. Morselli and F. Moura and K. Nakazawa and L. Nava and D. Nieto and M. Orienti and M. Orio and E. Orlando and P. Orleanski and S. Paiano and R. Paoletti and A. Papitto and M. Pasquato and B. Patricelli and M.Á. Pérez-García and M. Persic and G. Piano and A. Pichel and M. Pimenta and C. Pittori and T. Porter and J. Poutanen and E. Prandini and N. Prantzos and N. Produit and S. Profumo and F.S. Queiroz and S. Rainó and A. Raklev and M. Regis and I. Reichardt and Y. Rephaeli and J. Rico and W. Rodejohann and G. {Rodriguez Fernandez} and M. Roncadelli and L. Roso and A. Rovero and R. Ruffini and G. Sala and M.A. Sánchez-Conde and A. Santangelo and P. {Saz Parkinson} and T. Sbarrato and A. Shearer and R. Shellard and K. Short and T. Siegert and C. Siqueira and P. Spinelli and A. Stamerra and S. Starrfield and A. Strong and I. Strümke and F. Tavecchio and R. Taverna and T. Terzić and D.J. Thompson and O. Tibolla and D.F. Torres and R. Turolla and A. Ulyanov and A. Ursi and A. Vacchi and J. {Van den Abeele} and G. Vankova-Kirilovai and C. Venter and F. Verrecchia and P. Vincent and X. Wang and C. Weniger and X. Wu and G. Zaharijaš and L. Zampieri and S. Zane and S. Zimmer and A. Zoglauer},

}

\end{document}